\documentclass{article}
\usepackage{fullpage}
\usepackage{amsthm}
\usepackage{amsmath}
\usepackage{amsfonts}
\usepackage{amssymb}

\usepackage{algorithm}
\usepackage{algpseudocode}
\usepackage{easy-todo}
\usepackage{color}
\usepackage{afterpage}
\usepackage{subcaption}
\usepackage{graphicx}
\usepackage{epstopdf}
\usepackage{lineno}
\usepackage[all]{xy}
\usepackage{natbib}
\usepackage{setspace}
\usepackage{fullpage}
\usepackage{rotating}
\usepackage{longtable}
\usepackage{framed}
\usepackage{placeins}

\definecolor{lightgray}{gray}{0.95}

\allowdisplaybreaks

\newcommand{\authornote}[1]{}

\newcommand{\anonymiseForJETS}[1]{***}
\newcommand{\ignore}[1]{}
\newcommand{\pjs}[1]{\marginpar{\sc pjs}{\textcolor{blue}{#1}}}

\newtheorem{theorem}{Example}
\newenvironment{example}{\begin{theorem}}{\end{theorem}}

\newtheorem{mydef}{Definition}
\newenvironment{definition}{\begin{mydef}\rm}{\end{mydef}}

\newenvironment{ctabbing}%
    {\begin{center}\begin{minipage}{\textwidth}\begin{tabbing}}%
    {\end{tabbing}\end{minipage}\end{center}}
					
\begin{document}

\newcommand{\cand}{\mathcal{C}}
\newcommand{\gone}{\mathcal{E}}
\newcommand{\stand}{\mathcal{S}}
\newcommand{\election}{\mathcal{B}}
\newcommand{\voters}{V}

\newcommand{\plusplus}{+\!\!+}
\newcommand{\mss}{\{\!\!\{}
\newcommand{\mse}{\}\!\!\}}


\title{Towards Computing the Margin of Victory in STV Elections}
\author{Michelle Blom\footnote{Corresponding author: michelle.blom@unimelb.edu.au}, Peter J. Stuckey, Vanessa J. Teague\\ Department of Computing and Information Systems \\The University of Melbourne\\ Parkville, Australia}
\date{}
\maketitle

\begin{abstract}
The Single Transferable Vote (STV) is a system of preferential voting employed
in multi-seat elections. Each vote cast by a voter is a (potentially partial)
ranking over a set of candidates. No techniques currently exist for computing the
\textit{margin of victory} (MOV) in STV elections. The MOV is the smallest number of vote
manipulations (changes, additions, and deletions) required to bring about
a change in the set of elected candidates. Knowledge of the MOV of an election
gives greater insight into both how much time and money should be spent on the auditing of 
the election, and whether uncovered mistakes (such as ballot box losses) throw
the election result into doubt---requiring a costly repeat election---or can
be safely ignored. In this paper, we present algorithms for computing lower
and upper bounds on the MOV in STV elections.
In small instances, these algorithms are able to compute exact margins. 
\end{abstract}

\section{Introduction}\label{sec:Intro}

The Single Transferable Vote (STV) is a system of preferential voting employed in multi-seat elections. It is used to elect candidates to the
Australian Senate, in all elections in Malta, and in most elections in the Republic of Ireland
\citep{farr05}. No techniques currently exist for computing the smallest number of vote
manipulations (changes, additions, and deletions) required to bring about
a change in the set of elected candidates---the \textit{margin of victory}
(MOV). The
ability to compute this margin has significant value. In the 2013 election of
six candidates to Western Australia's Senate a discrepancy of 1,375
initially verified votes was discovered during a
recount. The election result was overturned, and a repeat election held in
2014. If the MOV for the original election was known, the question
of whether the loss of these votes may have altered the resulting outcome
could have been answered. In this instance, if the MOV was greater than 1,375 votes, the inclusion of these 1,375 lost votes would not have changed the election outcome. 


In an STV election, each vote is a (potentially partial) ranking over a set
of candidates. For example, in an election with candidates $c_1,$ $c_2,$ $c_3,$
and $c_4,$ a vote [$c_2,$ $c_1,$ $c_4$] expresses a first preference for
candidate $c_2$, a second for $c_1,$ and a third for $c_4$. At the start of the counting process, each vote is initially awarded to its highest ranked candidate. In the above vote, $c_2$
is the highest ranked candidate. 
The votes awarded to each candidate forms their
\textit{tally}.  Candidates whose tallies exceed (or reach) a \textit{quota}, defined in
terms of the number of seats to be filled and votes cast in the election, are elected
to a seat. As each candidate is elected,
their \textit{surplus} (the number of votes by which their tally exceeds
the quota) is computed, and a subset of their votes (with a combined value
equal to the surplus) is distributed to their next preferred candidate (in the above vote, the next preferred candidate after $c_2$ is $c_1$).
Where multiple candidates have a quota's worth of votes in their tally, the candidate
with the largest surplus is elected first, and their surplus is distributed. Then,
if there are still seats to fill, the candidate with the next largest surplus is
elected, and their surplus distributed (and so on).  
If no remaining candidate has a quota's worth of votes, and one or more seats remain empty,
the candidate with the fewest votes is eliminated and their votes distributed to their next preferred candidates.
If the number of candidates remaining (unelected and
not yet eliminated) equals the number of seats left to be filled, these
candidates are elected and the STV counting process terminates. 

Several STV variants exist, differing in the way that surpluses are distributed
\citep{weeks11}. 
Consider a candidate with a tally of 100 votes and a surplus of 40 votes. The
Inclusive Gregory Method redistributes all 100 votes, each with an assigned 
value of 0.4 (each vote is worth 0.4 votes), to their next highest ranked
candidate that is `still standing' (has not yet been elected or eliminated) \citep{mir04}.  In the STV
variant we consider in this paper, candidates whose tallies have
reached or exceeded the quota (but have not yet been awarded a seat) receive no further votes from the surplus
distributions of other candidates. 

Computing the MOV of an STV election is an extremely complex combinatorial optimisation problem. In an election with $n$ candidates, and $k$ available seats, there are ${n}\choose{k}$ possible allocations of candidates to these seats, and $n!$ different orders in which candidates can appear in the elimination and election sequence. Our task is to find an election outcome, out of the $n!$${n}\choose{k}$ possibilities, that differs from the original outcome of the election \textit{and} that requires the least number of vote manipulations to realise.   

We develop, in this paper, an algorithm for computing exact margins of victory in STV elections that use the Inclusive Gregory Method of surplus distribution---arguably the simplest and most straightforward of the existing
variants.  In Section
\ref{sec:STVCounting}, we step through the counting
process that takes place in STV elections, under the Inclusive Gregory Method,
in two example STV instances.  The algorithm we present in this paper (labelled {\bf{margin-stv}}) is an adaptation of existing work for computing margins in Instant Runoff Voting (IRV) elections \citep{blom16}. IRV is a single-seat variant of STV  and is employed in lower
house elections across Australia, in which a single candidate is elected to a
single seat. In an IRV election, candidates with the fewest votes are eliminated, and 
their votes redistributed to later ranked candidates, until a
candidate attains a majority of the available votes, and is declared the winner. The computation 
of exact IRV and STV victory margins is known to be NP-hard
\citep{bartholdi1991single,conitzer2007elections,Xia12,rothe13,wash14}.  

Our {\bf margin-stv} algorithm represents the outcome of an STV election as a sequence of candidate elections and eliminations (e.g., $c_4$ elected, $c_3$ eliminated, $c_2$ eliminated, $c_1$ elected). We present a mixed-integer non-linear program (MINLP) that, given such an outcome, and the set of votes cast in the election, computes the smallest number of vote manipulations required to realise the outcome. A vote manipulation replaces the ranking of a vote (e.g., [$c_2,$ $c_1,$ $c_4$]) with an alternate ranking (e.g., [$c_4,$ $c_3$]). Consider an election over candidates $\mathcal{C}$, in which candidates $E \subset \mathcal{C}$ are elected to a seat. Our {\bf margin-stv} algorithm applies branch-and-bound to search the space of alternate election outcomes (in which the set of winning candidates $E'$ $\neq$ $E$) for one that requires the least number of vote manipulations to realise.  
We show that {\bf margin-stv} is able to compute exact margins in some small STV election instances. We develop a relaxation of this algorithm capable of computing \textit{lower bounds} on the margin of victory in larger, more realistic STV election instances. 

\ignore{As in the work of \cite{Magrino:IRV} and \cite{blom15}, {\bf margin-stv} computes margins under the assumption that any manipulation
applied to cast votes must leave the number of
votes \textit{unchanged}. \cite{blom15} consider two variations of their algorithm for computing IRV margins in which this assumption is not required. The first variation assumes that votes can only be added to the election (and not removed, or changed). The second assumes that votes can only be removed from those cast (and not changed, or added). Computing the IRV MOV under these assumptions provides answers to important practical questions. If some people voted multiple times, could this have influenced the election outcome? Alternately, if some cast votes have been lost, could their inclusion have altered the election outcome? These types of manipulation are known as \textit{voter control}. Like \cite{blom15}, we describe how our {\bf margin-stv} algorithm can be adapted to compute STV margins under these two different assumptions. }

The remainder of this paper is structured as follows. Section \ref{sec:STVCounting} describes the STV counting algorithm. Preliminary definitions and concepts required in the explanation of our {\bf margin-stv} algorithm are presented in Section \ref{sec:Preliminaries}. Related work is discussed in Section \ref{sec:RelatedWork}.  Mechanisms for computing simple upper bounds on the degree of manipulation required to alter the outcome of an STV election are presented in Section \ref{sec:SimpleBounds}. Section \ref{sec:MINLP} presents our MINLP for computing the smallest degree of manipulation required to realise a specific election outcome (a specific sequence of elections and eliminations). A rule for computing a lower bound on this degree of manipulation is presented in Section \ref{sec:LowerBounds}. 
Using these upper and lower bounding techniques as building blocks, we present our {\bf margin-stv} algorithm, and a relaxed variant of this algorithm, in Section \ref{sec:MarginSTV}. We evaluate these algorithms on a range of both small and large STV election instances in Section \ref{sec:Evaluation}. 

\section{The Single Transferable Vote (STV)}\label{sec:STVCounting}

This section describes the STV vote counting algorithm that we consider in this paper,  
 outlined in Figure \ref{stvcount}. We illustrate this algorithm in the example election shown in Table \ref{tab:EGSTV1}.
In Step 1 of the counting algorithm the \textit{quota} of the election is calculated, according
to Equation \ref{eqn:Droop}. This is known as the \textit{Droop} quota, and
represents a \textit{threshold} that each candidate must reach before they are
elected to a seat. In the election of Table \ref{tab:EGSTV1} there are 2 seats to be filled and 60 cast votes. The quota in this election is 21 votes.

\begin{equation}
\text{Quota } = \frac{\text{Total number of votes cast}}{\text{Number of seats} + 1} +
1
\label{eqn:Droop}
\end{equation} 

Each vote cast in the election starts with a value of 1. The total value of the
votes a candidate has in their tally is computed as shown in Equation
\ref{eqn:totalvalue}. As the STV algorithm proceeds, votes will move from the
tally of one candidate to that of others. The value of these votes---the extent 
to which they contribute to a candidate's tally value---will change over the
course of the algorithm. 

\begin{equation}
\text{Total value of votes} = \sum_{\text{votes in tally}} \text{Vote value}
\label{eqn:totalvalue}
\end{equation}

Each vote is assigned to its first ranked candidate (Step
2). The first ranked candidate of  vote [$c_1,$ $c_4,$ $c_3$] is $c_1$. The
total value of the votes in each candidate's tally is computed. Table \ref{tab:EGSTV1b} shows that the 
tally values of candidates $c_1$ to $c_4$ after this first round of counting are 26, 10, 9, and 15.
The candidates whose tally value equals or exceeds the quota are placed in
a list sorted in decreasing order of \textit{surplus} size. A candidate's surplus is
equal to the difference between their tally value and the quota, as shown in
Equation \ref{eqn:surplus}. In our example, $c_1$ is the only candidate whose tally has exceeded the quota with a surplus of 5 votes.

\begin{equation}
\text{Surplus} = \text{Tally value } - \text{ Quota}
\label{eqn:surplus}
\end{equation}

If this list of candidates with a surplus (denoted \textit{surpluses} in Figure
\ref{stvcount}) is \textit{not} empty, we select the
first candidate in this list, $c,$ and elect them to a seat (Step 9). If we have filled all available seats, the algorithm terminates (Step 10). Otherwise, we 
redistribute the votes in $c$'s tally. Each vote is placed in the tally of
the next most preferred \textit{eligible} candidate after $c$ in its ranking.
Candidates that have been elected, eliminated, or whose tally value equals or
exceeds the quota are \textit{not eligible}. The set of votes in $c$'s tally
that have no eligible next preferred candidate are exhausted (not redistributed). The remainder are
labelled \textit{transferable} (Step 11).
These votes will have a \textit{reduced value} when they are redistributed. Their current value is reduced by a factor $\tau$ known as the \textit{transfer value}, computed (in Step 12) as shown in Equation
\ref{eqn:tvalue}.

\begin{equation}
\tau = \min \left(1, \frac{\text{Surplus of candidate } c}{\text{Total value of } c\text{'s
transferable votes}}\right)
\label{eqn:tvalue}
\end{equation}   

\begin{table}
    \begin{subtable}{.4\columnwidth}
      \centering
        \begin{tabular}{cr}
& \\
& \\
\hline
Ranking & Count \\
\hline
{}[$c_2$, $c_3$] & 4 \\
{}[$c_1$] &  20 \\
{}[$c_3$, $c_4$] & 9 \\
{}[$c_2$, $c_3$, $c_4$] & 6 \\
{}[$c_4$, $c_1$, $c_2$] &  15 \\
{}[$c_1$, $c_3$] & 6 \\
\hline
\end{tabular}
        \caption{}
				\label{tab:EGSTV1a}
    \end{subtable}$\,\,\,$
    \begin{subtable}{.4\columnwidth}
      \centering
        \begin{tabular}{crrr}
Seats: 2 & & & \\
Quota: 21  & & & \\
\hline
Candidate & Round 1 & Round 2 & Round 3 \\
\hline
 & $c_1$ elected & $c_2$ eliminated & $c_3$ elected\\
& $\tau_1$ $=$ $0.83$ & & \\
\hline
{}$c_1$ & 26  & --- & ---\\
{}$c_2$ &  10 & 10 & --- \\
{}$c_3$ & 9   & 14 & 24 \\
{}$c_4$ & 15  & 15 & 15 \\
\hline
\end{tabular}
\caption{}
\label{tab:EGSTV1b}
    \end{subtable} 
    \caption{Example 1: An STV election profile, stating (a) the number of
votes cast with each listed ranking over candidates $c_1$ to $c_4$, and (b)
the tallies after each round of counting, election, and elimination. }
\vspace{-0.4cm}
\label{tab:EGSTV1}
\end{table}

Consider our example in Table \ref{tab:EGSTV1}. Candidate $c_1$ is elected to a seat, and has 6 transferable votes with ranking [$c_1,$ $c_3$]. The remaining votes in $c_1$'s tally have a ranking of [$c_1$]. These votes have no eligible next preference and are exhausted (not redistributed). The transfer value assigned to $c_1$'s transferable votes is 0.83. All 6 votes with ranking [$c_1,$ $c_3$] are given to candidate $c_3$, but they now have a combined value of 5. 

\begin{figure}[t]
\begin{framed}
\begin{ctabbing}
xxxx \= xx \= xx \= xx \= xx \= xx \= xx \= xx \= xx \= xx \= \kill
1 \> Compute the \textit{Quota} for the election (Equation \ref{eqn:Droop}) \\
2 \> Assign votes to their first ranked candidates (each vote $b$ has a value of
$v_b = 1$)  \\[5pt] 
3 \> \textbf{While} a seat remains to be filled \textbf{do} \\
4 \>\>  Let \textit{surpluses} denote the candidates whose tally equals or \\
\>\>\> exceeds the quota (in order of decreasing \textit{surplus} size) \\[5pt]
5 \>\> \textbf{If} the \textit{surpluses} set is empty \textbf{then} \\
6 \>\>\> Eliminate a candidate $c$, of those remaining, with the smallest tally \\
7 \>\>\> Redistribute each vote $b$ in $c$'s tally to its
next preferred candidate at its current value $v_b$ \\[5pt]
8 \>\> \textbf{Else} \\
9 \>\>\> Elect the first candidate in \textit{surpluses}, $c$, to a seat \\[5pt]
10 \>\>\> \textbf{if} all seats are filled \textbf{then} stop \\[5pt]
11 \>\>\> Compute the number of \textit{transferable} votes in $c$'s tally
\\[2pt]
12 \>\>\> Compute the \textit{transfer value} $\tau$ of these \textit{transferable}
votes\\[2pt]
13 \>\>\> Redistribute each transferable vote $b$ in $c$'s tally with a value of
$\tau \, v_b$, where $v_b$ \\
\>\>\>\>  is its current value, to its next preferred candidate (skipping over candidates \\
\>\>\>\>  that have been elected, eliminated, or  are in the \textit{surpluses} set)\\[5pt]
14 \>\> \textbf{If} the number of unfilled seats and remaining candidates are
equal \textbf{then} \\
15 \>\>\> Elect all remaining candidates to a seat 
\end{ctabbing}
\vspace{-0.2cm}
\end{framed}
\vspace{-0.2cm}
\caption{The STV vote counting algorithm (under the Inclusive Gregory Method).}
\label{stvcount}
\end{figure}

If, in Step 4, no candidate has a tally value that equals or exceeds the
quota, the candidate $c$ with the smallest tally value is eliminated (Step 6). The votes in their tally are
redistributed to later preferences. The votes in $c$'s tally  that have $c$ as a
\textit{first} preference will have a value of 1 (as a consequence of Step 2). Votes that $c$ has received
after prior surplus distributions will have a reduced value. All votes in
$c$'s tally that have an eligible next preferred candidate (a candidate that is
still standing) are given to that candidate \textit{at their current value}
(Step 7). If the contribution to $c$'s tally value of a vote is $v$, then that
vote's contribution to its next preferred candidate's tally is also $v$. Consider again the example of Table \ref{tab:EGSTV1}. After the election of candidate $c_1$ and the redistribution of their votes, candidates $c_2$ to $c_4$ have tally values of 10, 14, and 15. No candidate has a quota of votes in their tally. Candidate $c_2$ has the smallest tally value, and is eliminated. Each of $c_2$'s votes---four with ranking [$c_2,$ $c_3$] and 6 with ranking [$c_2,$ $c_3,$ $c_4$]---still have a value of 1 and are redistributed to candidate $c_3$. Candidates $c_3$ and $c_4$ now have tally values of 24 and 15. 

After a candidate has been elected or eliminated, we check whether the number of
unfilled seats  and the number of remaining candidates (that have not yet
been elected or eliminated) are equal (Step 14). If so, all remaining candidates are
elected to a seat irrespective of their tally value (Step 15). If not, the total value of the votes in each remaining
candidate's tally is recomputed and we return to Step 4. In the example of Table \ref{tab:EGSTV1}, the algorithm recomputes the tally values of candidates $c_3$ and $c_4$ in Step 4, and places $c_3$ in the \textit{surpluses} list. Candidate $c_3$ is elected to the final seat, and the algorithm terminates in Step 10.
The STV  algorithm proceeds in rounds that consist of: computing the total value
of each candidate's tally; electing the candidate with the largest surplus (if
such a candidate exists) and redistributing their votes; or eliminating the
candidate with the smallest tally value (if no candidate has a quota)
and redistributing their votes. 

Let us consider a second example STV election, shown in Table \ref{tab:EGSTV2}. Candidates $c_1$, $c_2$, $c_3$, and $c_4$ have initial tallies of 31, 17, 5, and 10 votes. The quota for the election is 22, and $c_1$ is placed into the list of candidates with a surplus in Step 4. Candidate $c_1$ is elected to the first of two available seats in Step 9, and has 13 transferable votes in their tally (Step 11). The transfer value to be applied to those votes is 0.69 (its surplus of 9 divided by the number of transferable votes 13). In Step 13, candidate  $c_2$ is given 5 votes of ranking [$c_1,$ $c_3$], with the total value of these votes equal to 3.46. Candidate $c_4$ is given 8 votes of ranking [$c_1,$ $c_4,$ $c_2,$ $c_3$], with the total value of these votes equal to 5.54.  In the second round of counting, $c_2$ now has a total tally value of 20.46 votes and $c_4$ a total tally value of 15.54. No candidate has a quota, and candidate $c_3$, with the smallest tally, is eliminated (in Step 6).  Candidate $c_2$ is given 5 votes with ranking [$c_3,$ $c_2,$ $c_4$]. In the next round of counting, candidate $c_2$ has exceeded a quota and is elected to the last seat (Step 9).

\begin{table}
    \begin{subtable}{.4\columnwidth}
      \centering
        \begin{tabular}{cr}
& \\
& \\
\hline
Ranking & Count \\
\hline
{}[$c_1$, $c_2$, $c_3$] & 5 \\
{}[$c_1$] &  18 \\
{}[$c_4$, $c_3$] & 10 \\
{}[$c_3$, $c_2$, $c_4$] & 5 \\
{}[$c_2$, $c_4$, $c_3$] &  17 \\
{}[$c_1$, $c_4$, $c_2$, $c_3$] & 8 \\
\hline
\end{tabular}
        \caption{}
				\label{tab:EGSTV2a}
    \end{subtable}$\,\,\,$
    \begin{subtable}{.4\columnwidth}
      \centering
        \begin{tabular}{crrr}
Seats: 2 & & & \\
Quota: 22  & & & \\
\hline
Candidate & Round 1 & Round 2 & Round 3 \\
\hline
 & $c_1$ elected & $c_3$ eliminated & $c_2$ elected\\
& $\tau_1$ $=$ $0.69$ & & \\
\hline
{}$c_1$ & 31  & --- & ---\\
{}$c_2$ & 17 & 20.46 & 25.46 \\
{}$c_3$ & 5  & 5 & --- \\
{}$c_4$ & 10  & 15.54 & 15.54 \\
\hline
\end{tabular}
\caption{}
\label{tab:EGSTV2b}
    \end{subtable} 
    \caption{Example 2: An STV election profile, stating (a) the number of
votes cast with each listed ranking over candidates $c_1$ to $c_4$, and (b)
the tallies after each round of counting, election, and elimination. }
\label{tab:EGSTV2}
\end{table}

In many STV variants, the last bundle of votes
received by a candidate, at any point in the counting process, is known as their \textit{last parcel}.  In the Original Gregory Method,
votes in an elected candidate's last parcel (and no others) are transferred (at a fraction of
their value) during surplus distribution. The total value of the votes
transferred is equal to the candidate's surplus. Some jurisdictions do not
assign fractional values to distributed votes, but transfer a \textit{random
selection} of votes from a candidate's last parcel at their full value, the
total of which equals the candidate's surplus. Much of
the complexities involved in vote distribution across these variants are in
place to support easier and faster manual vote counting.  \cite{weeks11} and \cite{mir04}
provide a good summary of the range of STV variants used in practice.

\ignore{
In earlier work, we present an efficient algorithm for computing victory margins in Instant Runoff Voting (IRV) elections \citep{blom15}. An IRV election seeks to elect a \textit{single} winner from a field of candidates. The counting process in an IRV election proceeds in rounds of candidate elimination. In the first round, the candidate with the least votes in their starting tally is eliminated. These votes are distributed to their next preferred candidate. Votes with no next preferred candidate become exhausted. The elimination of the candidate with the smallest tally, and the redistribution of their votes, continues until a single candidate remains. This last remaining candidate is declared the winner. Unlike IRV, STV elects multiple candidates with the counting algorithm alternating between rounds of candidate \textit{election} and candidate \textit{elimination}. The votes in the tally of elected candidates are redistributed at a \textit{reduced value}. }

\section{Preliminaries}\label{sec:Preliminaries}

Definition \ref{def:election} formally defines an STV election
$\mathcal{E}$. Our representation of the outcome of an STV election---as a
sequence of candidate elections and eliminations---is outlined in
Definition \ref{def:Order}. The margin of victory (MOV) of an STV election
is defined in Definition \ref{def:MOV}.  The primary vote of a candidate $c$
is computed as described in Definition \ref{def:Pvote}, and used to compute
simple upper bounds on the MOV of an STV election in Section
\ref{sec:SimpleBounds}.

\begin{definition}\textbf{STV Election ($\mathcal{E}$)} An STV election is defined as a tuple  $\mathcal{E} = (\mathcal{C}, \mathcal{B}, \mathcal{Q}, N, E)$ where $\mathcal{C}$ is a set of candidates, $\mathcal{B}$ the set of votes cast in the election, $\mathcal{Q}$  the  election quota (the number of votes a candidate must attain to be elected to a seat, as defined in Equation \ref{eqn:Droop}), $N$ the number of seats to be filled, and $E$ the set of candidates elected to a seat (according to the counting algorithm outlined in Figure \ref{stvcount}). Each vote $b \in \mathcal{B}$ is a partial or complete ranking over  $\mathcal{C}$ (e.g., the vote $[c_1, c_3, c_2]$, in an election with candidates $\mathcal{C} = \{c_1, c_2, c_3, c_4\}$, expresses a first preference for candidate $c_1$, a second for $c_3$, and a third for $c_2$).
\label{def:election}
\end{definition} 

\begin{definition}\textbf{Election Order ($\pi$)} Given an STV election $\mathcal{E} = (\mathcal{C}, \mathcal{B}, \mathcal{Q}, N, E)$, we represent the outcome of the election as an election order $\pi$---a sequence of tuples $(c, a)$ where $c \in \mathcal{C}$ and $a \in \{0, 1\}$. The tuple  $(c,1)$ indicates that candidate $c$ is elected to a seat, and $(c,0)$ that $c$ is eliminated. An election order $\pi$ defines the sequence of elections and eliminations that arise as the STV counting algorithm (Figure \ref{stvcount}) is executed. The order $\pi$ $=$ $[(c_1,1),$ $(c_3,0),$ $(c_2,0),$ $(c_4,1)]$, for example, indicates that candidate $c_1$ is elected to a seat in the first round of counting, followed by the elimination of candidates $c_3$ and $c_2$, and the election of $c_4$.     
\label{def:Order}
\end{definition}

\begin{definition}\textbf{Margin of Victory (MOV)} The margin of victory for
  an STV election $\mathcal{E} = (\mathcal{C}, \mathcal{B}, \mathcal{Q}, N,
  E)$ is defined as the smallest number of vote \emph{manipulations}
  required to ensure that a set of candidates $E' \neq E$ are elected to a
  seat (i.e., at least one candidate in $E'$ must not appear in $E$).
  A single \emph{manipulation} changes the ranking on a single vote $b$ to an alternate ranking. For example, consider a vote $b$ with ranking [$c_1,$ $c_3,$ $c_2$]. Replacing $b$'s ranking with the alternate ranking [$c_4,$ $c_1$] represents a single manipulation. 
\label{def:MOV}   
\end{definition}

\begin{definition}\textbf{Primary Vote $f_p(c)$} The primary vote of a candidate $c \in \mathcal{C}$ in an STV election $\mathcal{E} = (\mathcal{C}, \mathcal{B}, \mathcal{Q}, N, E)$ is defined as the total number of votes in $\mathcal{B}$ in which $c$ is ranked highest (i.e., $c$ is ranked first). For example, the vote $[c_1, c_3, c_4]$ contributes to the primary vote of candidate $c_1$. 
\label{def:Pvote}
\end{definition}

\section{Related Work}\label{sec:RelatedWork}

The computation of victory margins in both STV and Instant Runoff Voting (IRV) elections is NP-hard \citep{bartholdi1991single,conitzer2003many,conitzer2007elections,Xia12,rothe13,wash14}. To the best of our knowledge, the algorithms presented in this paper form the first attempt to compute margins in STV elections. \cite{blom16} present a branch-and-bound algorithm for computing victory margins in IRV elections, itself an adaptation of earlier work by \cite{Magrino:IRV}. An IRV election elects a single winner $w$ from a field of candidates $\mathcal{C}$ on the basis of the votes $\mathcal{B}$   
cast in the election. As in an STV election, each vote $b \in \mathcal{B}$ is a (possibly partial) ordering over the candidates in $\mathcal{C}$. An IRV election proceeds in rounds of candidate elimination. Each vote is placed in the \textit{tally} of the candidate $c$ that appears \textit{first} in its ordering. In the first round, the candidate with the least votes in their tally is eliminated. Each of these votes is placed in the tally of its next preferred candidate. Votes with no next preferred candidate become exhausted (are not redistributed). This process of candidate elimination is repeated until only a single candidate remains---this candidate is the winner of the election. Unlike IRV, STV elects multiple candidates with the counting algorithm alternating between rounds of candidate \textit{election} and candidate \textit{elimination}. The votes in the tallies of elected candidates are redistributed at a \textit{reduced value}.   

\cite{Magrino:IRV} represent the outcome of an IRV election as a candidate sequence $\pi$ with candidates listed in the order in which they are eliminated (with the last candidate being the winner). Given such a sequence $\pi$, and a collection of votes, a linear program (LP) is presented that computes the smallest number of vote manipulations required to realise $\pi$. This linear program is labelled \textsc{DistanceTo}. In an election with winner $w$, \cite{Magrino:IRV} search the space of alternate elimination sequences (in which a candidate \textit{other than} $w$ is elected) for one requiring the least number of vote manipulations to realise. A key observation made by \cite{Magrino:IRV} is that, given a partial sequence of candidates $\pi'$, the \textsc{DistanceTo} LP computes a \textit{lower bound} on the number of vote manipulations required to realise any elimination order that \textit{ends in} $\pi'$. \cite{Magrino:IRV} progressively explore and build partial candidate elimination orders in a branch-and-bound algorithm. The \textit{last round margin} (LRM) of the election (defined as the difference in the tallies of the winning candidate and runner-up, divided by two and rounded up) is used as an upper bound on the MOV. Consider an IRV election with candidates $c_1$, $c_2$, $c_3$, and $c_4$, with outcome [$c_4,$ $c_3,$ $c_2,$ $c_1$], where $c_1$ is the winning candidate. Partial orders containing a single candidate (not including the original winner $c_1$) are added to a tree. \textsc{DistanceTo} is applied to each partial order $\pi'$ in this tree to compute a lower bound on the number of vote manipulations required to realise an elimination sequence ending in $\pi'$. The partial order $\pi'$ with the smallest \textsc{DistanceTo} evaluation is expanded by adding a candidate (not already in $\pi'$) to the start of the sequence. Partial orders with evaluations equal to or larger than the current upper bound are pruned. When a complete elimination order (involving all candidates) is formed, its \textsc{DistanceTo} evaluation is used to revise the current recorded upper bound. The algorithm terminates once all partial orders have either been expanded or pruned, with the revised upper bound returned as the  MOV.

\cite{blom16} improve the efficiency of the branch-and-bound algorithm of
\cite{Magrino:IRV} by introducing new rules for computing lower bounds on
the manipulation required to realise each partial order in the search
tree. These rules typically result in \textit{tighter} (i.e., higher) lower
bounds for each partial order than supplied by solving the
\textsc{DistanceTo} LP. Consequently, \cite{blom16} are able to: prune
larger portions of the space of partial elimination sequences; reduce the
number of calls to the \textsc{DistanceTo} LP; and quickly compute margins in
elections for which the algorithm of \cite{Magrino:IRV} times out after
72 hours of computation. Our {\bf margin-stv} algorithm shares a similar
structure to that of \cite{blom16} and \cite{Magrino:IRV}, in that it
searches the space of alternate election and elimination sequences using
branch-and-bound. The {\bf margin-stv} algorithm differs in several key
aspects: each node is a partial sequence of candidate elections and
eliminations (in place of a sequence of eliminations); a MINLP (in place of
an LP) is used to evaluate nodes in this search tree; and the descendants of
a partial sequence $\pi'$ are all complete sequences that \textit{start
  with} $\pi'$ (in place of all sequences than \textit{end in}
$\pi'$). Moreover, a variation of the winner elimination upper bound for IRV
elections \citep{cary:irv} is used as an initial upper bound on the STV MOV
(as described by \cite{chil16}).  Section \ref{sec:MarginSTV} describes the
\textbf{margin-stv} algorithm in detail.

The winner elimination upper bound (on the IRV margin of victory) of \cite{cary:irv} finds the most efficient way to eliminate the apparent winner of an IRV election at each elimination round, returning the least-cost (involving the smallest number of vote changes) of these. \cite{chil16} develop a version of this upper bound for use in STV elections. Figure \ref{alg:WEUB} outlines this STV variant of the winner elimination upper bound. Consider the example STV election of Table \ref{tab:EGSTV1}, where $c_1$ was elected in the first round of counting, $c_2$ eliminated in the second, and $c_3$ elected in the third (at which point both available seats had been filled). Following the algorithm listed in Figure \ref{alg:WEUB}, the winner elimination upper bound ($weub$) is initially set to the total number of votes cast in the election (Step 1), which is 60. The set of candidates that are (eventually) elected is $E$ $=$ \{$c_1,$ $c_3$\}. In the first round of counting, a candidate ($c_1$) is elected. The algorithm moves on to the second round (Step 3), in which a candidate ($c_2$) is eliminated with 10 votes ($v_2 = 10$ in Step 6). In Step 7, we consider each candidate in $E$ that has not been elected by round 2---candidate $c_3$---and determine how they could be eliminated in this round. Candidate $c_3$ has 14 votes in round 2 ($w_{2} = 14$ in Step 8). We could certainly eliminate $c_3$ by taking 4 votes from their tally ($\Delta = 4$ in Step 9) and giving them to some other candidate ($c_2$, for example). However, candidate $c_3$ can still be eliminated in this round if we take 2 votes from their tally and give them to $c_2$ (Steps 11 and 12), under the assumption that we can break the resulting tie between $c_3$ and $c_2$ in $c_2$'s favour. The winner elimination upper bound is set to 2 in Step 12. The algorithm moves on to the last round of counting in which no candidate is eliminated.  Steps 5 to 12 are skipped and a winner elimination upper bound of 2 is returned.    
   
\begin{figure}[t]
\begin{framed}
\begin{ctabbing}
xxxx \= xx \= xx \= xx \= xx \= xx \= xx \= xx \= xx \= xx \= \kill
1 $weub$ $\leftarrow$ $|\mathcal{B}|$ \\
2 $E$ $\leftarrow$ candidates (eventually) elected \\[5pt]
3 \textbf{for each} round of counting $j$ \textbf{do} \\
4 \> \textbf{if} a candidate is eliminated in round $j$ \textbf{do} \\
5 \>\> $c_j$ $\leftarrow$ candidate eliminated in round $j$ \\
6 \>\> $v_j$ $\leftarrow$ number of votes in $c_j$'s tally in round $j$ \\[5pt]
7 \>\> \textbf{for each} $w \in E$ that has not yet been elected by round $j$ \textbf{do} \\
8 \>\>\> $w_j$ $\leftarrow$ number of votes in $w$'s tally in round $j$ \\
9 \>\>\> $\Delta$ $\leftarrow$ $\lceil w_j - v_j \rceil$ \\
10 \>\>\> $weub$ $\leftarrow$ $\min(\Delta, weub)$\\[5pt]
11 \>\>\> \textbf{if} $\lceil w_j - \frac{1}{2} v_j \rceil$ is less than or equal to tallies of all candidates still standing (excluding $c_j$) \textbf{do}\\
12 \>\>\>\> $weub$ $\leftarrow$ $\min(\lceil w_j - \frac{1}{2} v_j \rceil, weub)$\\
13 \textbf{return} $weub$
\end{ctabbing}
\vspace{-0.3cm}
\end{framed}
\vspace{-0.2cm}
\caption{The winner elimination upper bound of \cite{cary:irv} applied to compute an upper bound (denoted $weub$) on the MOV of an STV election (as used by \cite{chil16}). The notation $\mathcal{B}$ and $E$ denote the set of votes cast in the election, and the set of candidates elected to a seat, respectively.}
\label{alg:WEUB}
\end{figure}

To the best of the authors knowledge, the work of \cite{chil16} describes the only attempt to compute bounds on the MOV for STV elections (in its adaptation of the winner elimination upper bound of \cite{cary:irv}). The algorithms we present in this paper are, to the best of our knowledge, the first attempts to compute exact margins in small STV elections and lower bounds on the MOV in larger instances. IRV elections, in contrast, have received more consideration. \cite{blom16} and \cite{Magrino:IRV} present algorithms for computing exact margins in IRV elections. A number of works have presented algorithms for computing lower and upper bounds on IRV margins (see \cite{cary:irv} and \cite{sarwate-checkoway-shacham:irv-audit:spp13}).

The focus of this paper is the computation of the MOV for STV elections. This MOV is defined as the smallest number of vote manipulations (the replacing of the ranking of a vote with an alternate ranking) required to ensure that a different set of candidates is elected (i.e., at least one candidate in this set is replaced with one that was not originally elected to a seat). Similar questions have been considered for alternate
voting rules. The complexity of manipulating an election 
with bribery is considered by \cite{falis09}, under a number of voting schemes:
Condorcet-based; approval
voting; scoring rules; veto rules; and plurality. Their aim is to
find a manipulation to achieve a desired election result, while minimising the
cost of bribes given to voters for changing their vote.  If the cost of bribing a voter to change their vote is 1, this least cost set of bribes is equivalent to the margin of victory, as we have defined it.  \cite{kaczmarczyk2016algorithms} consider a variant of the bribery problem (destructive shift bribery) in which voters can be bribed to demote the position of a candidate $c$ in their ranking by $p$ positions (i.e., to move a candidate down in their ranking by $p$ positions) in a bid to ensure that $c$ does not win the election. The authors analyse the complexity of destructive shift bribery for a number of voting rules ($k$-Approval, Borda, Copeland and Maximin). Polynomial-time algorithms are presented for computing the smallest set of desired bribes in the case of the k-Approval, Borda, and Maximin rules, while the problem is shown to be NP-complete for the Copeland rule. \cite{schurmann2017automatic} describe a model-checking-based approach for the computation of margins in D'Hondt elections, applying their approach to the 2015 Danish national parliamentary elections.  The STV elections we consider in this paper vary considerably from the voting rules considered by \cite{kaczmarczyk2016algorithms}, and  \cite{schurmann2017automatic}. We refer the reader to \cite{brandt2016handbook} for further discussion on bribery and manipulation problems in the computational social choice context.  

Given an STV election $\mathcal{E}$, our {\bf margin-stv} algorithm relies on: a mechanism for computing an upper bound on the manipulation required to change the outcome of $\mathcal{E}$; a mechanism for computing the smallest manipulation required to realise a specific outcome for $\mathcal{E}$---a specific sequence of candidate eliminations and elections $\pi$; and a mechanism for computing a lower bound on the degree of manipulation required to realise a candidate order that \textit{starts} with a partial sequence of elections and eliminations $\pi'$. These mechanisms are provided in Sections \ref{sec:SimpleBounds} to \ref{sec:LowerBounds}, and form the basis of our branch-and-bound algorithm.

\section{Simple Upper Bounds on the STV MOV}\label{sec:SimpleBounds}

Figure \ref{alg:WEUB} presents an algorithm for computing an upper bound (a winner elimination upper bound) on the STV MOV\footnote{This algorithm has been extracted from the work of \cite{chil16}. The algorithm is not explicitly presented in that paper, but can be found in code, implemented by Andrew Conway, for analysing the 2012 local government elections in New South Wales, Australia. This code is available at: https:/github.com/SiliconEconometrics/PublicService}.  In elections where all seats are filled by a candidate prior to any eliminations taking place (e.g., an election with 2 seats, 4 candidates, and outcome [$(c_1,1)$ $(c_2,1),$ $(c_3, 0),$ $(c_4, 0)$]), this algorithm is not able to reduce the upper bound from its original value (the total number of votes cast in the election). In these instances, we introduce a simple bound on the STV MOV, computed as follows. Consider an election $\mathcal{E} = (\mathcal{C}, \mathcal{B}, \mathcal{Q}, N, E)$. The STV counting algorithm of Figure \ref{stvcount} elects candidates to a seat once the number of votes in their tally reaches or exceeds $\mathcal{Q}$. To change the outcome of $\mathcal{E}$, with winning candidates $E \subset \cand$, we must find a series of vote manipulations that elects a candidate $c \in \cand \setminus E$ to a seat. It is clear that we can elect  $c \in \cand \setminus E$, with primary vote $f_p(c)$ (the total number of votes in $\mathcal{B}$ in which $c$ is ranked \textit{first}), if we take $\mathcal{Q} - f_p(c)$ votes away from other candidates and give them to $c$ (we replace the ranking of these votes by a ranking that preferences $c$ first).  We compute $\mathcal{Q}- f_p(c)$ for each $c \in \cand \setminus E$ and take the smallest result as an upper bound on the STV MOV.  We call this bound the Simple-STV upper bound.

\section{Computing Minimal Manipulations: A MINLP}\label{sec:MINLP}

Given an IRV election and a sequence of candidate eliminations $\pi'$, \cite{Magrino:IRV} present a linear program (LP) for computing the smallest number of vote manipulations required to ensure candidates are eliminated in the order specified in $\pi'$. In this section, we present a mixed-integer non-linear program (MINLP) that, given an STV election $\mathcal{E} = (\mathcal{C}, \mathcal{B}, \mathcal{Q}, N, E)$, and a candidate order $\pi$ (a sequence of candidate elections and eliminations), computes the smallest number of vote manipulations (changes to votes in $\mathcal{B}$) required to realise $\pi$. We re-use some notation and constraints introduced in the LP of \cite{Magrino:IRV}. The votes in $\mathcal{B}$ form the \textit{original profile} of the election. The MINLP below, denoted \textsc{DistanceTo}$_{STV}$, introduces variables indicating which votes in $\mathcal{B}$ are to be changed, and what their ranking will be in a \textit{new} or \textit{modified profile}. The constraints of the following MINLP are designed to enforce a specific candidate order $\pi$ by modifying the smallest number of votes in $\mathcal{B}$ (where required). 

\subsection{Notation}\label{sec:Notation}

\noindent\textbf{Sets and Indices} 
\begin{longtable}{p{20pt}p{400pt}}
$s, \mathbf{S}$ & A signature $s \in \mathbf{S}$ is a partial or total ranking over the candidates in $\mathcal{C}$; $\mathbf{S}$ is the set of \textit{all possible} partial or total rankings over $\mathcal{C}$ (including those that do not appear on a vote in $\mathcal{B}$). \\[3pt]
$\mathcal{R}$ & The set of rounds of counting in the election---in each round a
candidate is either elected to a seat or eliminated. \\[3pt]
$\mathcal{P}_{ij}$ & The subset of vote signatures that could
possibly be in the tally of candidate $i \in \mathcal{C}$ at the start of round
$j$ (this can be inferred on the basis of the order of candidates in $\pi$).\\[3pt]
$\pi[j]$ & The candidate that is elected or eliminated in round $j$ (according to $\pi$).\\[3pt]
$\mathcal{B}_{sij}$ & The set of candidates that appear \textit{before} candidate $i$
in the preference ordering of signature $s \in
\mathbf{S}$ that are \textit{still standing} (have not been elected or eliminated) at the end of round $j$. \\[3pt]
 $\mathcal{S}_j$ & The set of candidates still standing \textit{at the start} of round $j$. \\[3pt]
$\pi^+$ & The subset of candidates in $\mathcal{C}$ that are elected to a seat according to $\pi$. \\[3pt]
$\pi^-$ & The subset of candidates in $\mathcal{C}$ that are eliminated according to $\pi$. 
\end{longtable}
\addtocounter{table}{-1}

\noindent\textbf{Constants}
\begin{longtable}{p{20pt}p{400pt}}
$N_s$ & Number of votes with signature $s \in \mathbf{S}$ cast in election $\mathcal{E}$ (i.e., the number of votes in $\mathcal{B}$ whose ranking matches signature $s$).\\[3pt]
$UB$ & Known upper bound on the number of vote manipulations required to
realise $\pi$ (such as the winner elimination upper bound of Figure \ref{alg:WEUB} or the Simple-STV upper bound of Section \ref{sec:SimpleBounds}). \\
\end{longtable}
\addtocounter{table}{-1}

\noindent\textbf{Decision Variables}
\begin{longtable}{p{20pt}p{400pt}}
$p_s$ & Integer number of votes in $\mathcal{B}$ that are modified so that their
new signature is $s \in \mathbf{S}$. \\[3pt]
$m_s$ & Integer number of votes whose signature in $\mathcal{B}$ (the original profile) is $s \in
\mathbf{S}$, but are modified to something other than $s$ in the new profile.
\\[3pt]
$y_s$ & Integer number of votes with signature $s \in \mathbf{S}$  in the new
election profile.\\[3pt]
$\tau_j$ & Transfer value of votes being redistributed from an elected
candidate in round $j$. \\[3pt]
$\rho_j$ & Number of votes eligible for transfer, from the candidate elected
to a seat in round $j$, to candidates that are still standing at the end of round $j$. \\[3pt]
$q_{ij}$ & Binary variable with a value of 1 if the tally of candidate $i$
exceeds or equals the quota at the start of round $j$, and 0 otherwise. \\[3pt]
$v_{ij}$ & Floating point number of votes in the tally of candidate $i$ at the start of round
$j$. \\[3pt]
$y_{ijs}$ & Floating point number of votes with signature $s \in \mathbf{S}$ in the tally of
candidate $i$ at the start of round $j$. \\[3pt]
$d_{ijs}$ & Floating point number of votes with signature $s \in \mathbf{S}$
transferred to candidate $i$ from the candidate elected or eliminated in round
$j$. \\
\end{longtable}
\addtocounter{table}{-1}

\subsection{Objective}
The objective of \textsc{DistanceTo}$_{STV}$ is to minimise the number of votes in $\mathcal{B}$ whose signature is changed. 
\begin{equation}
z = \min \sum_{s \in \mathbf{S}} p_s 
\end{equation}

\subsection{Constraints}
Constraints \eqref{eqn:balance} and \eqref{eqn:ys} appear in the LP of \cite{Magrino:IRV}. Constraint \eqref{eqn:balance} ensures that the total number of votes in the new election profile (after manipulation) is equal to the total number of votes in the original profile. This assumption is made by both \citet{blom16} and \citet{Magrino:IRV}. \citet{blom16} explore variations of the \textsc{DistanceTo} LP of \citet{Magrino:IRV} in which this constraint is removed, and a manipulation may \textit{add} votes (for example, arising from voters voting multiple times), or remove votes (for example, arising from ballot box losses). We intend, in future work, to consider the `deletion-only' and `addition-only' settings analysed by \citet{blom16} for IRV elections, in the STV setting. Constraint \eqref{eqn:ys} defines the number of votes with signature $s \in \mathbf{S}$ in the new profile, $y_s$. 
  
\begin{flalign}
\sum_{s \in \mathbf{S}} m_s & = \sum_{s \in \mathbf{S}} p_s & & \label{eqn:balance} \\
y_s &  = N_s + p_s - m_s  & & \forall~s \in \mathbf{S} \label{eqn:ys}
\end{flalign}

Constraint \eqref{eqn:quotaflow} ensures that candidates who have achieved a quota by round $j$, have also achieved a quota by round $j+1$. Constraints \eqref{eqn:havequota} and \eqref{eqn:noquota} ensure that that number of votes a candidate $i$ has in their tally at the start of round $j$ is greater than or equal to $\mathcal{Q}$ if $i$ has achieved a quota by round $j$ ($q_{ij}$ is 1) and less than $\mathcal{Q}$ otherwise ($q_{ij}$ is 0, and $\epsilon \ll 1$). 

\begin{flalign}
q_{i,j-1} & \leq  q_{ij} & & \forall~i \in \mathcal{C}, j \in \mathcal{R} \label{eqn:quotaflow}\\
v_{ij} & \geq q_{ij} \, \mathcal{Q} & & \forall~i \in \mathcal{C}, j \in \mathcal{R} \label{eqn:havequota}\\
v_{ij} & \leq (1 - q_{ij}) \, (\mathcal{Q} - \epsilon) + q_{ij} \, |\mathcal{B}| & & \forall~i \in \mathcal{C}, j \in \mathcal{R} \label{eqn:noquota} 
\end{flalign}

Constraints \eqref{eqn:yij1} and \eqref{eqn:yij2} define the number of votes
of signature $s \in \mathbf{S}$ in candidate $i$'s tally at the start of
round $j$. Where $j > 1$, this is equal to the number of votes of signature
$s \in \mathbf{S}$ in candidate $i$'s tally at the start of the previous
round, plus the number of votes of signature $s \in \mathbf{S}$ distributed
to $i$ from the candidate elected (or eliminated) in round $j - 1$. As all
votes of a single signature $s$ will reside in the tally of only one
candidate at any point in time, at least one of $y_{i,j-1,s}$ and
$d_{i,j-1,s}$ will be zero in each instance of Constraint
\eqref{eqn:yij2}. Although the order in which candidates are elected and
eliminated is fixed (to that defined by $\pi$), the round in which a
candidate is given, via distribution, votes of certain signatures can
vary. Consider a candidate $c$ whose tally has reached or exceeded the quota
in round $j$. Candidate $c$ may have to wait several rounds to be elected to
a seat, if one or more other candidates have also reached a quota by round
$j$, and have more votes in their tallies (they are elected before
$c$). While candidate $c$ is waiting to be elected, they are not given any
additional votes during the distribution of the surpluses of elected
candidates---these votes skip $c$ and are given to the next eligible
candidate in their ranking. As the candidate order $\pi$ does not prescribe
exactly when an elected candidate achieves a quota, we must support the
possibility that votes of certain signatures $s \in \mathbf{S}$ can be
distributed to a candidate in one of a number of different rounds.
Constraint \eqref{eqn:vij} sums the total number of votes for candidate $i$
at the start of round $j$.

\begin{flalign}
y_{i1s} & = y_s & & \forall~i \in \mathcal{C}, s \in \mathcal{P}_{ij} \label{eqn:yij1}\\
y_{ijs} & = y_{i,j-1,s} + d_{i,j-1,s} & & \forall~i \in \mathcal{C}, j \in \mathcal{R}, j > 1, s \in \mathcal{P}_{ij} \label{eqn:yij2}\\
v_{ij} &= \sum_{s \in \mathcal{P}_{ij}} y_{ijs} & & \forall~i \in \mathcal{C}, j \in \mathcal{R} \label{eqn:vij}
\end{flalign}

Constraint \eqref{eqn:hasmorevotes} ensures that the candidate elected to a seat in round $j$ has more (or an equal number of) votes in their tally at the start of round $j$ than all other candidates who are still standing (have not been eliminated or elected before round $j$). Conversely, a candidate who is eliminated in round $j$ must have fewer (or an equal number of) votes in their tally at the start of round $j$ than all other candidates who are still standing (Constraint \eqref{eqn:haslessvotes}). Moreover, a candidate can only be eliminated in a round if no other remaining candidate has a quota's worth of votes in their tally at the start of the round (Constraint \eqref{eqn:elimnoquota}).  
\begin{flalign}
v_{\pi[j],j} & \geq v_{kj} & & \forall~j \in \mathcal{R} . \pi[j] \in \pi^+, k \in \mathcal{S}_j \setminus \{\pi[j]\} \label{eqn:hasmorevotes}\\
v_{\pi[j],j} & \leq v_{kj} & & \forall~j \in \mathcal{R} . \pi[j] \in \pi^-, k \in \mathcal{S}_j \setminus \{\pi[j]\} \label{eqn:haslessvotes} \\
q_{ij} & = 0  & & \forall~j \in \mathcal{R} . \pi[j] \in \pi^-, i \in \mathcal{S}_j \label{eqn:elimnoquota}
\end{flalign}

Constraint \eqref{eqn:define-d} defines the value of variable $d_{ijs}$ (the total number of votes of signature $s$ distributed to candidate $i$ in round $j$ after candidate $\pi[j]$ is elected or eliminated). Recall that the set $\mathcal{P}_{ij}$ denotes the subset of signatures $s \in \mathbf{S}$ that could possibly be in candidate $i$'s tally at the start of round $j$. The set $\mathcal{P}_{\pi[j],j} \cap \mathcal{P}_{i,j+1}$ contains only those signatures that could possibly be distributed from $\pi[j]$ (the candidate elected or eliminated in round $j$) to candidate $i$---the subset of signatures that could have been in $\pi[j]$'s tally in round $j$ \textit{and} in the tally of candidate $i$ at the start of round $j+1$. If $\pi[j]$ is eliminated ($\pi[j] \in \pi^-$), any votes distributed to another candidate are distributed at their current value ($d_{ijs} = y_{\pi[j],j,s}$ for all $s  \in \mathcal{P}_{\pi[j],j} \cap \mathcal{P}_{i,j+1}$). If $\pi[j]$ is elected, candidate $i$ may receive their votes of signature $s \in \mathcal{P}_{\pi[j],j} \cap \mathcal{P}_{i,j+1}$ (at a reduced value) only if $i$ does not already have a quota's worth of votes and is either the next preferred candidate in $s$, or all  candidates that appear between $\pi[j]$ and $i$ in $s$ already have a quota's worth of votes.   

\begin{flalign}
 d_{ijs}  & =
\begin{cases}
y_{\pi[j],j,s} & \quad \text{if } \pi[j] \in \pi^- \\
\tau_j \, y_{\pi[j],j,s} \,\,(1 - q_{ij}) \prod\limits_{k \in \mathcal{B}_{sij}} q_{kj} & \quad \text{if } \pi[j] \in \pi^+ 
\end{cases} & & \forall~i \in \mathcal{C}, j \in \mathcal{R}, s \in \mathcal{P}_{\pi[j],j} \cap \mathcal{P}_{i,j+1} \label{eqn:define-d}
\end{flalign}

Upon the election of a candidate $c$, a subset of the votes in their tally (those for which there is a next preferred candidate that is still standing---not yet eliminated or elected---and whose tally has not already reached or exceeded the quota) are distributed to one or more alternate candidates. These votes are called \textit{transferable votes}. The remainder become exhausted (there is no such `next preferred' candidate for these votes). The number of transferable votes in a candidate $c$'s tally, upon their election in round $j$, denoted $\rho_j$, is defined in Constraint \eqref{eqn:define-tvotes}. The transfer value assigned to these votes is dependent on both the size of $c$'s surplus (the value of votes in their tally minus the quota) and the total value of the transferable votes in $c$'s tally (the votes that have a valid next preferred candidate). This transfer value $\tau_j$, and its relationship to the quantity of transferable votes in an elected candidate's tally $\rho_j$, is defined in Constraint \eqref{eqn:define-tv}.  Both of these variables are relevant only in rounds in which a candidate is elected to a seat, and a surplus is distributed. 

\begin{flalign}
\rho_j & = \sum_{i \in \mathcal{S}_j} \sum_{s \in \mathcal{P}_{\pi[j],j} \cap \mathcal{P}_{i,j+1}}  y_{\pi[j],j,s}\,\, (1 - q_{ij}) \prod\limits_{k \in \mathcal{B}_{sij}} q_{kj} & & \forall~j \in \mathcal{R}, \pi[j] \in \pi^+ \label{eqn:define-tvotes}\\
\tau_j \, \rho_{j} & = v_{\pi[j],j} - \mathcal{Q} & & \forall~j \in \mathcal{R}, \pi[j] \in \pi^+ \label{eqn:define-tv} 
\end{flalign}

Constraint \eqref{eqn:define-tv} yields incorrect transfer values---that are greater than one---in the event that the total value of the transferable votes in an elected candidate $c$'s tally \textit{is less than} their surplus. This can occur if a large portion of $c$'s votes become exhausted once they are elected (votes that define a partial ordering over the set of candidates become exhausted when the \textit{last} candidate in their ordering is eliminated or elected). In this circumstance, the transfer value of $c$'s transferable votes is set to 1 (i.e., these votes are distributed at their current value). This is consistent with how STV elections are counted in practice. Constraint \eqref{eqn:define-tv} is thus rewritten as shown in Constraint \eqref{eqn:define-tv2}.

\begin{flalign}
\tau_j & =  
\begin{cases}
1 & \quad \text{if } \rho_j \leq v_{\pi[j],j} - \mathcal{Q} \\
v_{\pi[j],j} - \mathcal{Q} & \quad \text{otherwise}
\end{cases} & & \forall~j \in \mathcal{R}, \pi[j] \in \pi^+ \label{eqn:define-tv2}
\end{flalign}  

\section{Computing Lower Bounds for Partial Candidate Orders}\label{sec:LowerBounds}

\cite{Magrino:IRV} present an LP for computing the smallest number of vote manipulations required to realise a sequence of candidate eliminations in a given IRV election. This LP, when applied to a partial elimination sequence  $\pi'$ (under the assumption that all candidates not in $\pi'$ have already been eliminated and their votes distributed to candidates in $\pi'$), computes a \textit{lower bound} on the manipulations required to realise an elimination sequence \textit{ending in} $\pi'$.  Similarly, we can apply \textsc{DistanceTo}$_{STV}$ to a \textit{partial} sequence of candidate elections and eliminations $\pi'$ (a partial candidate order)  to compute a lower bound on the manipulations required to realise a complete order (including all candidates) that \textit{starts with} $\pi'$. 

Consider the STV election of Table \ref{tab:EGSTV1} with candidates $c_1,$ $c_2,$ $c_3,$ and $c_4$. Given the partial order [$(c_1, 0),$ $(c_3, 1)$], \textsc{DistanceTo}$_{STV}$   computes the smallest number of vote manipulations required to ensure that:
\begin{itemize}
\setlength\itemsep{0pt}
\item Candidate $c_1$ has the fewest votes in the 1$^{st}$ round of counting (fewer votes than $c_2$, $c_3$, and $c_4$);
\item No candidate has a quota's worth of votes in their tally in the 1$^{st}$ round;
\item Candidate $c_3$ has a quota's worth of votes in their tally in the 2$^{nd}$ round; and,
\item Candidate $c_3$ has the most votes in their tally in the 2$^{nd}$ round (more votes than $c_2$ and $c_4$).
\end{itemize}  
For any complete order $\pi$ that starts with [$(c_1, 0),$ $(c_3, 1)$], \textsc{DistanceTo}$_{STV}$  will ensure that the above constraints hold in addition to constraints that enforce the remaining elections and eliminations in $\pi$. For any partial order $\pi'$, the set of constraints enforced by \textsc{DistanceTo}$_{STV}$ is a subset of those enforced for any complete order starting with $\pi'$. Consequently, solving \textsc{DistanceTo}$_{STV}$ for $\pi'$ yields a lower bound on the manipulations required to realise any complete order starting with $\pi'$.  This property of the \textsc{DistanceTo}$_{STV}$ MINLP allows us to form a branch-and-bound algorithm for computing the MOV in STV elections.     
 
\subsection{A Simple Lower Bounding Rule}\label{sec:SimpleLowerBound}
\cite{blom16} present two methods for computing lower bounds on the degree of manipulation required to realise IRV elimination sequences that end in a partial order $\pi'$, without requiring the solving of an LP. We adapt the logic underlying these rules to develop a lower bounding rule applicable to partial orders in STV elections. Given a partial order $\pi'$ (a sequence of candidate eliminations and elections), this rule computes a lower bound on the number of vote manipulations required to realise any complete order starting with $\pi'$. 

Consider an STV election $\mathcal{E} = (\mathcal{C}, \mathcal{B}, \mathcal{Q}, N, E)$, and a partial order $\pi'$. 
We can infer from $\pi'$, the set of vote signatures $s \in \mathbf{S}$ that could potentially lie in the tally of each candidate $c \in \mathcal{C}$ in each round of election or elimination $j$ in $\pi'$ (this set is denoted $\mathcal{P}_{cj}$, as defined in Section \ref{sec:Notation}). Consequently, we can infer the maximum possible number of votes that could lie in the tally of each candidate $c$ in each round  $j$ (the total number of votes cast with a signature $s \in \mathcal{P}_{cj}$), $V^{max}_{cj}$, under the assumption that no manipulation has yet taken place (Equation \ref{eqn:MaxVotes}). Recall that $N_s$ denotes the number of votes in $\mathcal{B}$ that have been cast with signature $s \in \mathbf{S}$. Irrespective of when a candidate $c$ is elected or eliminated, $c$'s tally will contain \textit{at least} all cast votes in which they are the first preference (their primary vote $f_p(c)$, as per Definition \ref{def:Pvote}). The minimum number of votes in $c$'s tally, in any round $j$, $V^{min}_{c}$, is equal to $c$'s primary vote (Equation \ref{eqn:MinVotes}).   

\begin{flalign}
V^{max}_{cj} & = \sum_{s \in \mathcal{P}_{cj}} N_s  & &  \forall c \in \mathcal{C}, j \in \{1, \ldots, |\pi'|\}\label{eqn:MaxVotes} \\
V^{min}_{c} & = f_p(c) & &  \forall c \in \mathcal{C} \label{eqn:MinVotes}
\end{flalign} 

We consider each election and elimination in $\pi'$. Let $c_j$ denote the candidate elected or eliminated in round $j$ of $\pi'$. If $c_j$ is elected, $c_j$ must have a quota's worth of votes in their tally (unless the number of seats left to be filled in round $j$ equals the number of candidates still standing at the start of round $j$). To ensure that $c_j$ has a quota's worth of votes at the start of round $j$, we must modify at least $l_q$ votes (Equation \ref{eqn:LQ1}). 
\begin{flalign}
l_q(c_j, \pi') & = \max(0, \mathcal{Q} - V^{max}_{c_jj}) & & \label{eqn:LQ1}
\end{flalign}
 
If candidate $c_j$ is eliminated, $c_j$ must have fewer votes than all other candidates still standing at the start of round $j$, denoted $\mathcal{S}_j$, and no remaining candidate can have a quota's worth of votes. To ensure that $c_j$ has fewer votes than all candidates in $\mathcal{S}_j$, we must modify at least $l^1_e$ votes (Equation \ref{eqn:LE1}). To ensure that no remaining candidate has a quota's worth of votes at the start of round $j$, we must modify at least $l^2_e$ votes (Equation \ref{eqn:LE2}). To ensure that $c_j$ is eliminated in round $j$, we must modify at least $l_e = \max(l^1_e, l^2_e)$ votes.
\begin{flalign}
l^1_e(c_j, \pi') & = arg\,max_{c' \in \mathcal{S}_j \setminus \{c_j\}} \max(0, V^{min}_{c_j} - V^{max}_{c'j}) & & \label{eqn:LE1} \\
l^2_e(c_j, \pi') & = arg\,max_{c' \in \mathcal{S}_j} \max(0, V^{min}_{c'} - \mathcal{Q}) & & \label{eqn:LE2}
\end{flalign} 
Equation \ref{eqn:LE2} will ensure that no candidate has \textit{more} than a quota's worth of votes in round $j$. To ensure that a candidate has \textit{less} than a quota, their tally value must be less than or equal to $\mathcal{Q} - \epsilon$, where $\epsilon \ll 1$. As we are computing a lower bound on required vote manipulation, we ignore the $\epsilon$ term for simplicity.   

A lower bound on the degree of manipulation required to realise a complete order starting with $\pi'$ is computed by taking the maximum of $l_q$ and $l_e$ for each candidate elected, and eliminated, in $\pi'$.
\begin{flalign}
lb(\pi') & = arg\,max_j 
\begin{cases}
l_q(c_j, \pi') & \text{if } c_j \text{ is elected} \\
l_e(c_j, \pi') & \text{if } c_j \text{ is eliminated}
\end{cases} & & 
\end{flalign}

\begin{example} Consider the STV election of Table \ref{tab:EGSTV1} and the candidate order $\pi$ $=$ [($c_3,$ 0), ($c_1,$ 1), ($c_2,$ 1), ($c_4,$ 0)], where  candidate $c_3$ is eliminated in Round 1, $c_1$ is elected in Round 2, and $c_2$ is elected Round 3. We now use our lower bounding rule to compute a lower bound on the number of vote manipulations required to realise this order of elections and eliminations. The original winners in this election are candidates $c_1$ and $c_3$ ($E = \{c_1, c_3\}$). Given the cast votes listed in Table \ref{tab:EGSTV1a}, we determine where these vote signatures would lie (i.e., in which candidates tally), in each round of counting (until all seats have been filled), assuming $\pi$ is realised (Table \ref{tab:EG1_voteslieA}). Prior to any manipulation of the cast votes, the maximum value of each candidates tally, in each round, is listed in Table \ref{tab:EG1_voteslieB}. 
\begin{table}
\centering
    \begin{subtable}{\columnwidth}
      \centering
\begin{tabular}{|c|c|c|c|}
\hline
Candidate & Round 1 & Round 2 & Round 3  \\
\hline
$c_1$ & [$c_1$], [$c_1,$ $c_3$]               &  [$c_1$], [$c_1,$ $c_3$]               &  --- \\
$c_2$ & [$c_2,$ $c_3$], [$c_2,$ $c_3,$ $c_4$] &  [$c_2,$ $c_3$], [$c_2,$ $c_3,$ $c_4$] &  [$c_2,$ $c_3$], [$c_2,$ $c_3,$ $c_4$] \\
$c_3$ & [$c_3,$ $c_4$]                        & ---& --- \\
$c_4$ & [$c_4,$ $c_1,$ $c_2$]                 & [$c_4,$ $c_1,$ $c_2$], [$c_3,$ $c_4$]  &  [$c_4,$ $c_1,$ $c_2$], [$c_3,$ $c_4$] \\
\hline
\end{tabular}
\caption{}
\label{tab:EG1_voteslieA}
\end{subtable}\\
  \begin{subtable}{\columnwidth}
      \centering
\begin{tabular}{|c|c|c|c|}
\hline
Candidate & Round 1 & Round 2 & Round 3  \\
\hline
$c_1$ & 26 & 26 &  --- \\
$c_2$ & 10 & 10 &  10 \\
$c_3$ & 9  & --- & --- \\
$c_4$ & 15 & 24 &  24 \\
\hline
\end{tabular}
\caption{}
\label{tab:EG1_voteslieB}
\end{subtable}
\caption{(a) Possible distribution of vote signatures, in each round of counting of the STV election of Table \ref{tab:EGSTV1}, assuming the candidate order $\pi$ $=$ [($c_3,$ 0), ($c_1,$ 1), ($c_2,$ 1), ($c_4,$ 0)] is realised; and (b) the maximum tally value for each candidate, in each round of counting, assuming no vote manipulation has yet taken place.} 
\label{tab:voteslie}
\end{table}  
 The minimum value of each candidates tally, in each round, is equal to their primary vote. Here, $V^{min}_{c_1} = 26$, $V^{min}_{c_2} = 10$, $V^{min}_{c_3} = 9$, and $V^{min}_{c_4} = 15$. We consider each election and elimination in $\pi$. To eliminate $c_3$ in Round 1, $c_3$ must have fewer votes than all other candidates. Here, $l^1_e(c_3, \pi) = 0$ (no manipulation is required to ensure $c_3$ has the fewest votes). No candidate can have a quota's worth of votes in Round 1. In this example, the quota is 21 votes, $l^2_e(c_3, \pi) = 5$, and $l_e(c_3,\pi) = 5$. To elect $c_1$ in Round 2, we must ensure that $c_1$ has a quota in Round 2. Here, $l_q(c_1, \pi) = \max(0, \mathcal{Q} - V^{max}_{c_12}) = \max(0, 21 - 26) = 0$.  To elect $c_2$ in Round 3, we must ensure that $c_2$ has a quota in Round 3. Here, $l_q(c_2, \pi) = \max(0, \mathcal{Q} - V^{max}_{c_23}) = \max(0, 21 - 10) = 11$. Consequently, a lower bound on the manipulation required to realise $\pi$ is $lb(\pi)$ $=$ $\max\{l_e(c_3, \pi),$ $l_q(c_1, \pi),$ $l_q(c_2, \pi)\}$ $=$ $\max\{5, 0, 11\}$ $=$ $11$. 
\label{eg:CompLB1}
\end{example}

We describe, in Section \ref{sec:MarginSTV}, how this lower bounding rule can be used to avoid solving the \textsc{DistanceTo}$_{STV}$ MINLP for some partial orders explored by the {\bf margin-stv} algorithm.
 
\section{Computing Margins in STV Elections (margin-stv)} \label{sec:MarginSTV}

\begin{figure}[!t]
\begin{framed}
\small
\begin{ctabbing}
xxx \= xx \= xx \= xx \= xx \= xx \= xxxxxxxxxxxxxxxxxx \= \kill
\textbf{margin-stv}($\mathcal{E} = (\mathcal{C}, \mathcal{B}, \mathcal{Q}, N, E)$) \\
1 \> $F$ $\leftarrow$ $\emptyset$ \\
2 \> $UB$ $\leftarrow$ \textbf{estimate-upper-bound}($\mathcal{E}$) \>\>\>\>\>\> $\triangleright$ Compute winner elimination or Simple-STV upper bound\\
3 \> $\mathcal{A}$ $\leftarrow$ \{$1,$ $0$\} \\
4 \> \textbf{for}($c \in \mathcal{C}$) \\
5 \>\> \textbf{for}($a \in \mathcal{A}$) \\
6 \> \>\> $\pi'$ $\leftarrow$ $[(c,a)]$ \\
7 \> \>\> $l$ $\leftarrow$ \textsc{DistanceTo}$_{STV}$($\mathcal{E}$, $\pi'$) \\ 
8 \> \>\> \textbf{if}($l < UB$) \\
9 \> \> \>\> $F$ $\leftarrow$ $F \cup \{(l,\pi')\}$  \\[2pt]
10 \> \textbf{while} $F \neq \emptyset$ \\
11 \> \> $(l,\pi')$ $\leftarrow$ $\arg \min F$  \>\>\>\>\> $\triangleright$ Select partial order with the smallest assigned lower bound \\[2pt]
12 \> \> $F$ $\leftarrow$ $F \setminus \{(l,\pi')\}$ \\[2pt]
13 \>  \> $UB$ $\leftarrow$ \textbf{expand}($\pi',$ $UB,$ $F,$ $\mathcal{E},$ $\mathcal{A}$) \\[2pt]
14 \> \> $F$ $\leftarrow$ \textbf{prune}($F,$ $UB$) \>\>\>\>\> $\triangleright$ Partial orders with lower bounds $\geq UB$ are pruned from $F$\\
15 \> \textbf{return} $UB$ \\
\\
\textbf{expand}($\pi',$ $UB,$ $F,$ $\mathcal{E} = (\mathcal{C}, \mathcal{B}, \mathcal{Q}, N, E),$ $\mathcal{A}$)  \\
16 \> $UB'$ $\leftarrow$ $UB$\\
17 \> \textbf{for}($c \in $ \textbf{standing}($\pi',$ $\mathcal{C}$)) \>\>\>\>\>\> $\triangleright$ For each candidate $c$ still standing at the end of $\pi'$\\
18 \> \> \textbf{for}($a \in \mathcal{A}$) \\
19 \> \>\> $\pi$ $\leftarrow$ $\pi' \plusplus [(c, a)]$ \>\>\>\> $\triangleright$ A new partial order is created by eliminating or electing $c$\\[2pt]
20 \> \>\> \textbf{if}(\textbf{valid}($\pi,$ $\mathcal{E}$)) \>\>\>\> $\triangleright$ Valid orders do not re-elect all candidates in $E$\\
21 \>\>\>\> $l$ $\leftarrow$ \textsc{DistanceTo}$_{STV}$($\mathcal{E}$, $\pi$)\\[2pt]
22 \>\>\>\> \textbf{if}($|\pi|$ $=$ $|\mathcal{C}|$) \\
23 \>\>\>\> \> $UB'$ $\leftarrow$ $\min$ \{$UB',$ $l$\} \\
24 \>\>\>\> \> \textbf{continue}\\[2pt]
25 \> \>\>\> \textbf{if}($l < UB'$) \\
26 \> \>\>\> \> $F$ $\leftarrow$ $F \cup \{(l,\pi)\}$ \>\> $\triangleright$ The new partial order is added to the frontier $F$ \\[2pt]
27 \> \textbf{return} $UB'$ 
\end{ctabbing}
\end{framed}
\caption{ The {\bf margin-stv} algorithm for computing the MOV of an STV election $\mathcal{E} = (\mathcal{C}, \mathcal{B}, \mathcal{Q}, N, E)$ with candidates $\mathcal{C}$, votes $\mathcal{B}$, quota $\mathcal{Q}$, seats $N$, and winners $E \subset \mathcal{C}$.}
\label{alg:MarginSTV}
\end{figure}

Figure \ref{alg:MarginSTV} defines our {\bf margin-stv} algorithm for computing the MOV of an STV election. The {\bf margin-stv} algorithm maintains an initially empty frontier of partial candidate orders $F$ (Step 1). An upper bound $UB$ on the STV MOV is computed in Step 2. In our implementation of {\bf margin-stv} we use the minimum of the winner elimination upper bound of Figure \ref{alg:WEUB} and the Simple-STV upper bound of Section \ref{sec:SimpleBounds}. The frontier $F$ is populated with partial candidate orders of size one (the election or elimination of a single candidate)---each of which is assigned a lower bound on required manipulations with \textsc{DistanceTo}$_{STV}$---in Steps 4 -- 9. Partial orders with a lower bound less than the current upper bound $UB$ are added to $F$ in Step 9.

\begin{figure}[!t]
\centering
\begin{framed}
\includegraphics[width=14cm]{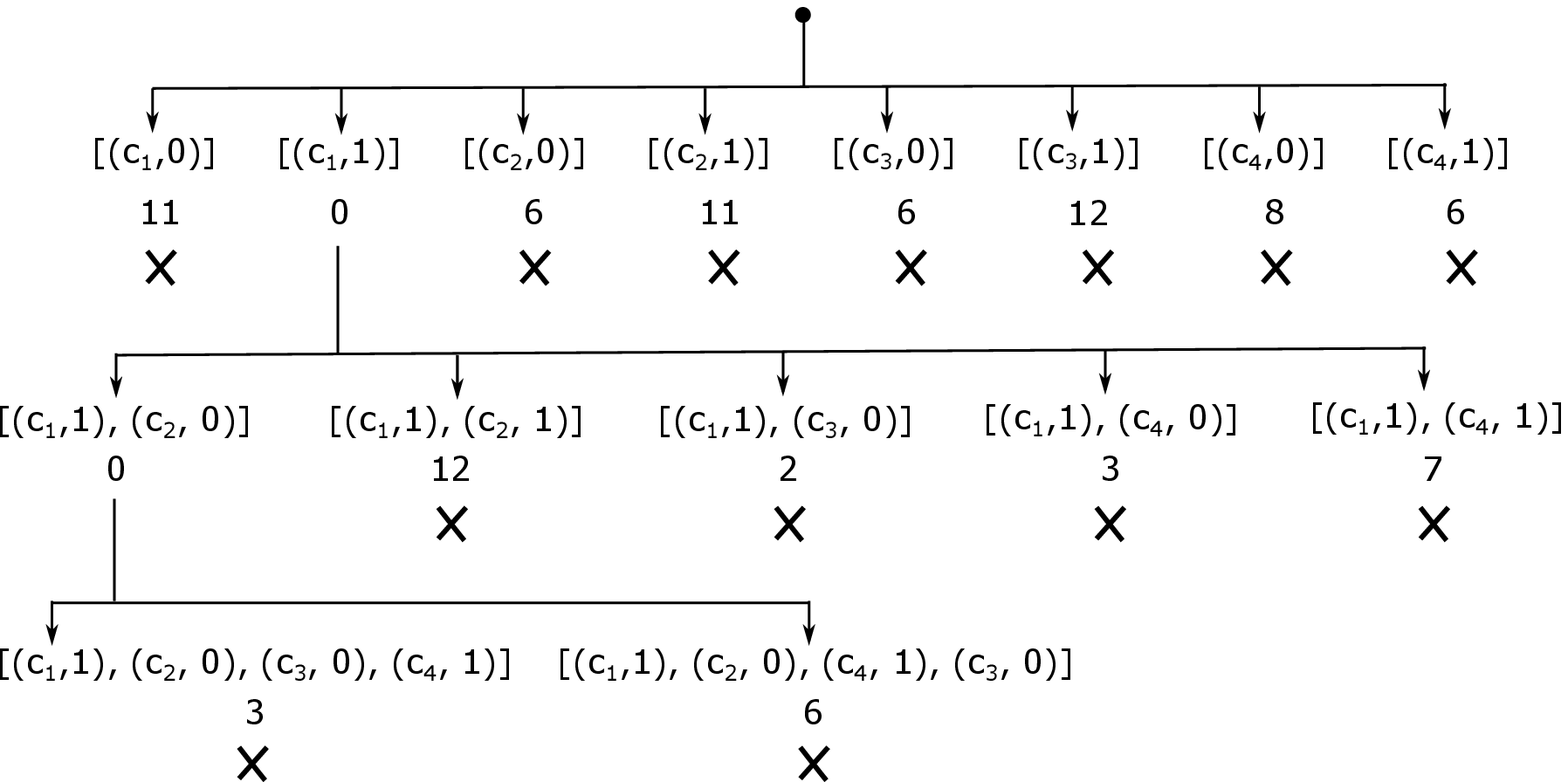}
\end{framed}
\caption{Complete search tree explored in the application of {\bf margin-stv} to the STV election of Table \ref{tab:EGSTV1}, among candidates $c_1,$ $c_2,$ $c_3,$ and $c_4$. The initial upper bound on the MOV in this example is 2 votes.}
\label{fig:traceEG1}
\end{figure}

\begin{example}
Consider the STV election of Table \ref{tab:EGSTV1}. The winner elimination upper bound  is 2 votes and the Simple-STV bound is 6 votes. In Step 2, the current upper bound $UB$ is initialised to 2. The partial candidate orders $[(c_1, 0)],$ $[(c_1, 1)],$ $[(c_2, 0)],$ $[(c_2, 1)],$ $[(c_3, 0)],$ $[(c_3, 1)],$ $[(c_4, 0)],$ and $[(c_4, 1)]$ are 
created, and assigned lower bounds by \textsc{DistanceTo}$_{STV}$ of 11, 0, 6, 11, 6, 12, 8, and 6, respectively. Only $[(c_1, 1)]$ is added to $F$ in Step 9. Figure \ref{fig:traceEG1} visualises the search tree explored by {\bf margin-stv} in this example.
\label{eg:StepAlg1}
\end{example}

Once the frontier $F$ is populated with partial orders of size one, Steps 10 to 14 expand these orders in the search for an alternate outcome for the election that requires the smallest degree of manipulation to realise. Partial orders with the smallest assigned lower bound are expanded in turn. A partial order $\pi'$ is expanded by first removing it from the frontier (Step 12). For each candidate $c$ not mentioned in $\pi'$, two new partial orders are formed in which $c$ is elected ($\pi' \plusplus [(c, 1)]$) and eliminated ($\pi' \plusplus [(c, 0)]$) in successive iterations of the loop in Steps 18 to 26. If the new partial order $\pi$ is \textit{valid} (i.e., it does not elect the same candidates elected in the original winners set $E$), it is evaluated with \textsc{DistanceTo}$_{STV}$ (Step 21). If $\pi$ contains all candidates (it is a complete order), it represents an alternate election outcome requiring $l$ vote manipulations to realise ($l$ is computed by \textsc{DistanceTo}$_{STV}$ in Step 21). If $l$ is less than the current upper bound, $UB$, $UB$ is replaced with $l$ (Step 23). If $\pi$ is not a complete order, it is added to $F$ if $l$ is less than $UB$ (Step 26). 

We repeatedly expand the partial order with the smallest assigned lower bound, updating the frontier $F$, until $F$ is empty (there are no remaining partial orders that can be expanded). At this point, the algorithm terminates and returns the current upper bound $UB$ as the election MOV in Step 15. 

\begin{example}[Example \ref{eg:StepAlg1} cont] The frontier, $F$, now contains one partial order $\pi'$ $=$ $[(c_1, 1)]$.  The following partial orders are formed and evaluated with \textsc{DistanceTo}$_{STV}$ in successive executions of Step 19 and 21:

\begin{longtable}{p{80pt}p{350pt}}
$[(c_1, 1), (c_2, 0)]$ & Lower bound of 0 \\[2pt]
$[(c_1, 1), (c_2, 1)]$ & Lower bound of 12 \\[2pt]
$[(c_1, 1), (c_3, 0)]$ & Lower bound of 2 \\[2pt]
$[(c_1, 1), (c_3, 1)]$ & Invalid as the original winners $c_1$ and $c_3$ are both elected \\[2pt]
$[(c_1, 1), (c_4, 0)]$ & Lower bound of 3 \\[2pt]
$[(c_1, 1), (c_4, 1)]$ & Lower bound of 7 \\
\end{longtable}
\addtocounter{table}{-1}

Only $[(c_1, 1),$ $(c_2, 0)]$ is added to $F$ in Step 26. At this point, $F$
$=$ $\{[(c_1, 1)$, $(c_2, 0)]\}$, and $[(c_1, 1)$, $(c_2, 0)]$ is the next
order expanded in Step 13. The following complete orders are formed in
successive executions of Step 19 (in our implementation of {\bf margin-stv},
we complete all partial orders formed in Step 19 with their `obvious' ending
if possible---i.e., if there is only one candidate remaining). All orders
have a lower bound greater than the current upper bound of 2, and are pruned
from consideration.

\begin{longtable}{p{140pt}p{290pt}}
$[(c_1, 1), (c_2, 0), (c_3, 0), (c_4, 1)]$ &  \textsc{DistanceTo}$_{STV}$ evaluation of 3\\[2pt]
$[(c_1, 1), (c_2, 0), (c_3, 1), (c_4, 0)]$  & Invalid as original winners $c_1$ and $c_3$ are elected \\[2pt]
$[(c_1, 1), (c_2, 0), (c_4, 0), (c_3, 1)]$ & Invalid as original winners $c_1$ and $c_3$ are elected \\[2pt]
$[(c_1, 1), (c_2, 0), (c_4, 1), (c_3, 0)]$ & \textsc{DistanceTo}$_{STV}$ evaluation of 6 \\
\end{longtable}
 \addtocounter{table}{-1}
 After pruning all partial orders from $F$, the frontier is empty and the algorithm  returns the upper bound of 2 as the election MOV in Step 15. In this example, {\bf margin-stv} has shown that there is no better manipulation (requiring fewer vote changes) than that inferred by the winner elimination upper bound.  
\label{eg:StepAlg2}
\end{example}

In Steps 7 and 21, where \textsc{DistanceTo}$_{STV}$ is evaluated for a partial (or complete) order $\pi$, we can use the lower bounding rule of Section \ref{sec:SimpleLowerBound} to compute an initial lower bound for $\pi$, $l'$. If $l' \ge UB$, the current upper bound, we need not solve \textsc{DistanceTo}$_{STV}$ as $\pi$ can immediately be pruned from consideration. If $l' < UB$, we can solve \textsc{DistanceTo}$_{STV}$ to get a second lower bound for $\pi$, $l$. We can then assign to $\pi$ the maximum of the two lower bounds $l'$ and $l$. In Section \ref{sec:Evaluation}, we examine the relative performance of {\bf margin-stv}, in terms of the number of \textsc{DistanceTo}$_{STV}$ MINLPs solved and the runtime of the algorithm, in the setting where this extra lower bounding computation is performed, and when it is omitted.

\subsection{Implementation Details}\label{sec:ImplementationDetails}

We use SCIP \citep{Achterberg2009} to solve each MINLP formed by {\bf
  margin-stv}. In practice, we have found that some instances of
\textsc{DistanceTo}$_{STV}$, even for small election instances, can be
difficult for SCIP to solve in reasonable time. Consequently, we
terminate MINLP solves if the time since a last improving
solution has been found
reaches a pre-specified time limit. If a partial order $\pi'$ is being evaluated, and
the MINLP is terminated before an optimal solution is found, the best
objective value (a lower bound on the optimal objective) is assigned to
$\pi'$ as its lower bound. If a complete order $\pi$ is being evaluated, and
the MINLP is terminated before finding an optimal solution, $\pi$ is
inserted into the frontier $F$ with a lower bound equal to the best
objective value of the MINLP, and the current upper bound \textit{is not}
revised (as would normally occur after evaluating a complete order). The
result is that {\bf margin-stv} may terminate with a frontier that contains
a number of complete orders (that cannot be further expanded) with a
smallest lower bound $L$, and a current upper bound $UB$ that is
\textit{greater than} $L$ (i.e., the algorithm returns with a lower and
upper bound on the MOV, but not an exact value). This occurs for a number of
elections in our test set in Section \ref{sec:Evaluation}.

\begin{figure}
  \begin{tabular}{c@{~~~~}c}
    \includegraphics[width=0.48\textwidth]{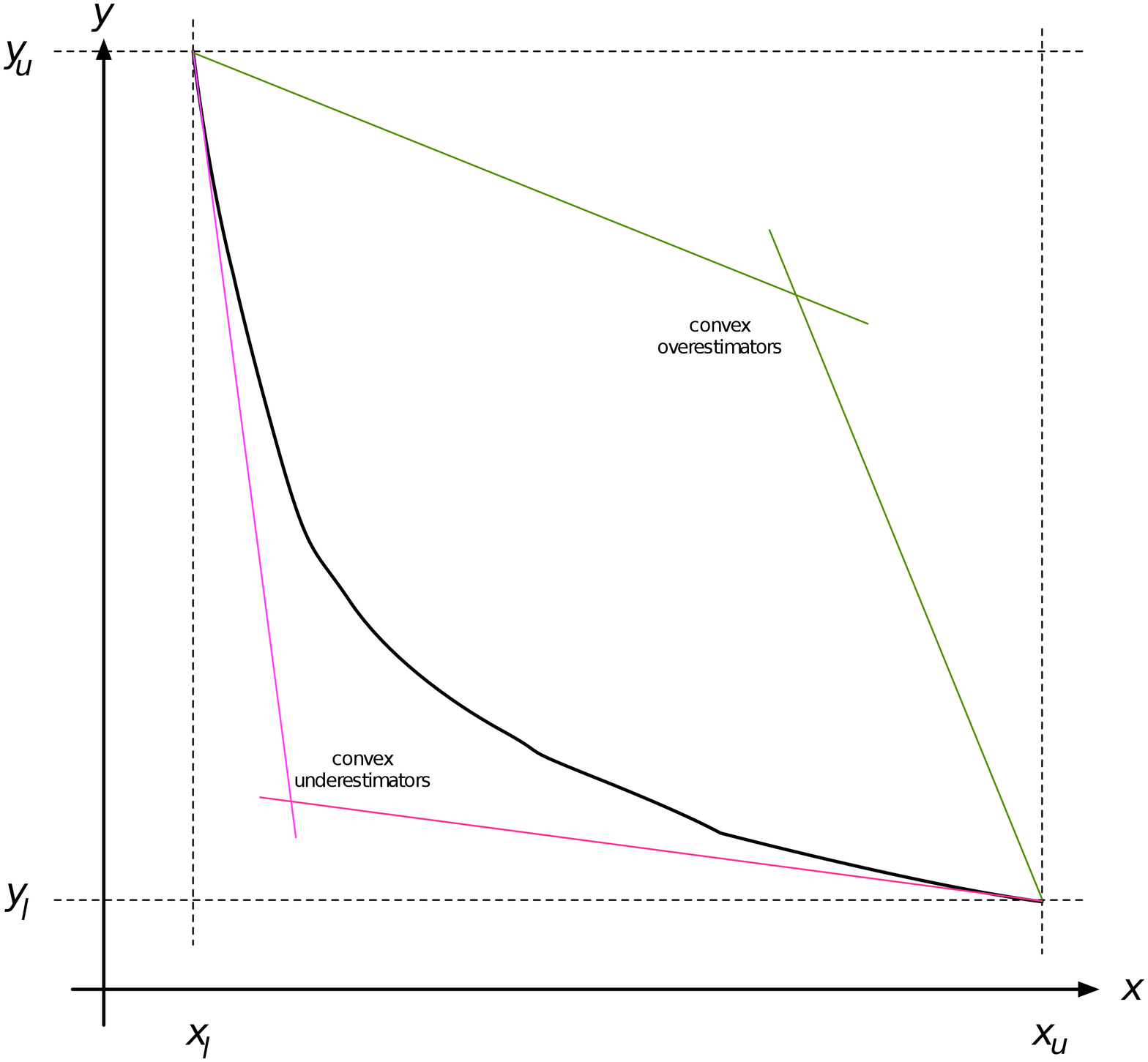}
  &
    \includegraphics[width=0.48\textwidth]{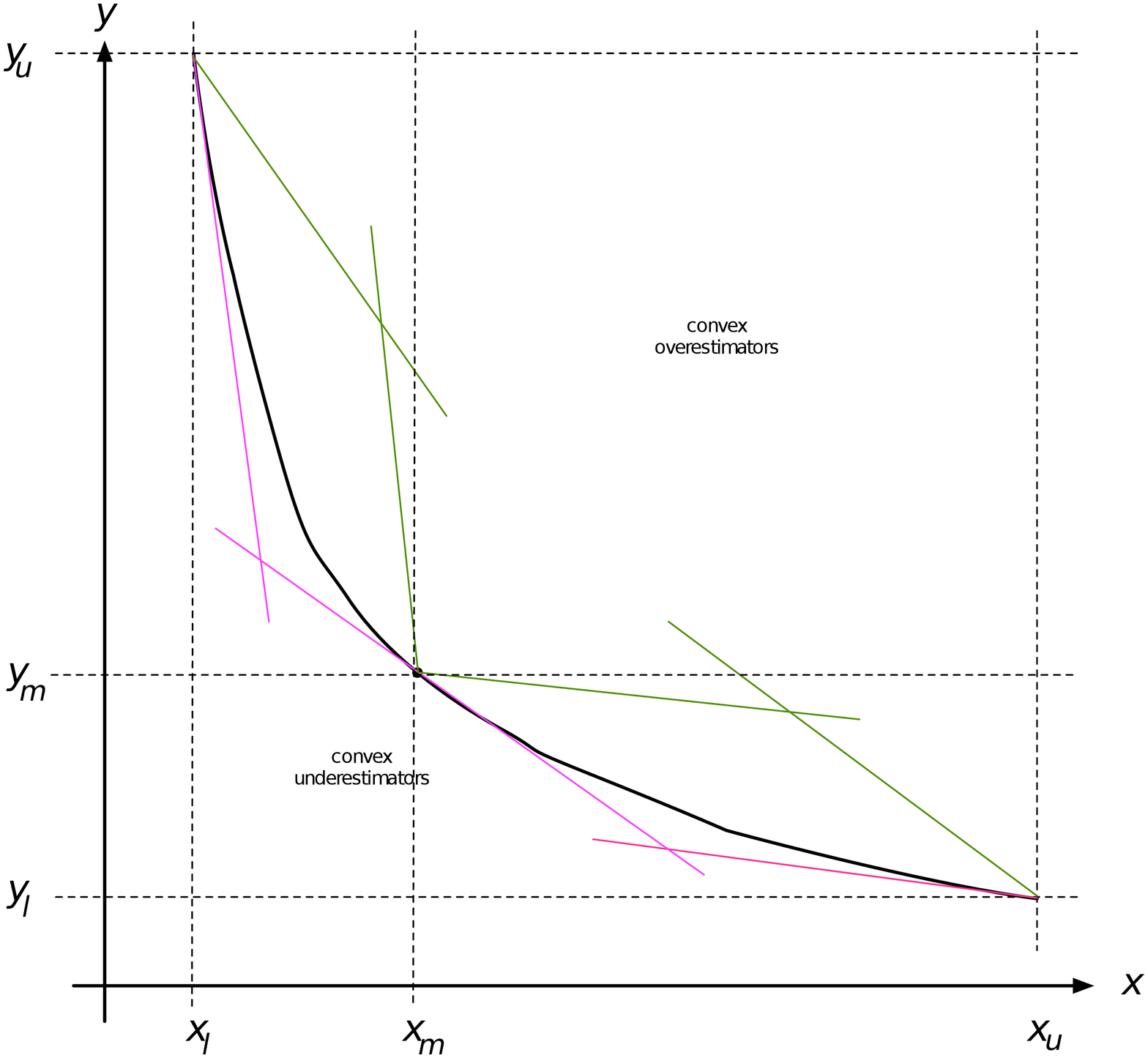}
    \\
    (a) & (b)
  \end{tabular}
  \caption{Illustration of McCormick envelopes (a) for $z = xy$ in the range
    $x \in x_l .. x_u$, $y \in y_l .. y_u$, and (b) when the range of $x$ is divided in two.}
  \label{fig:mccormick}
\end{figure}

It is clear from the evaluation of {\bf margin-stv}, in Section
\ref{sec:Evaluation}, on small STV elections (with candidate numbers ranging
from 4 to 13) that it will not scale to more realistic elections with dozens
of candidates. To improve its scalability, we vary the algorithm so that a
MIP relaxation of the \textsc{DistanceTo}$_{STV}$ is constructed to evaluate
each partial and complete order formed in Steps 6 and 19 of Figure
\ref{alg:MarginSTV}. We consider two types of relaxation. The first replaces
each bilinear term present in the MINLP (these terms appear in Constraints
\eqref{eqn:define-d} and \eqref{eqn:define-tv2}) with \cite{mcc76}
inequalities. Each bilinear term $x \, y$, where $x$ and $y$ are continuous
variables with domains $[x^L, x^U]$ and $[y^L, y^U]$, is replaced with
variable $z = x \, y$. Variable $z$ is defined by Equations \eqref{eqn:MC1}
to \eqref{eqn:MC2}. The optimal solution of the relaxed
\textsc{DistanceTo}$_{STV}$ is a \textit{lower bound} on that of the
MINLP. The MOV computed by {\bf margin-stv}, in this context, is a lower
bound on the true MOV.
An illustration of the McCormick envelope is shown in
Figure~\ref{fig:mccormick}(a)
when $z$ is fixed. 

\begin{flalign}
& z \geq y^L \, x + x^L \, y - x^L \, y^L & & \label{eqn:MC1}\\
& z \geq y^U \, x + x^U \, y - x^U \, y^U & & \\
& z \leq y^U \, x + x^L \, y - x^L \, y^U & & \\
& z \leq y^L \, x + x^U \, y - z^U \, y^L & & \label{eqn:MC2} 
\end{flalign}

Our second relaxation replaces each bilinear term present in the MINLP (these terms appear in Constraints \eqref{eqn:define-d} and \eqref{eqn:define-tv2}) with a piecewise linear relaxation (specifically, a relaxation proposed by \cite{gou09}). The result is again a MIP relaxation of the \textsc{DistanceTo}$_{STV}$ whose optimal solution is a \textit{lower bound} on that of the MINLP.  
 Given a product of two continuous variables $x$ and $y$ (a bilinear term
 $x\, y$), the piecewise linear relaxation we apply is defined as follows
 (replicated from the work of \cite{gou09}), where the variable $z = x\, y$
 replaces the bilinear term where it appears in the original MINLP. We
 divide the domain of $x$ into $K$ ranges ($[x_0, x_1],$ $\ldots,$
 $[x_{k-1}, x_k]$) and introduce a binary variable $\lambda_k$ to indicate
 whether $x$ has a value in range $[x_{k-1}, x_k]$. Constraint
 \eqref{eqn:SOS1} ensures that only one $\lambda_k$ can take on a value of
 1. The domain of variable $y$ is $[y^L,y^U]$. The variable $\delta y_k$
 defines the ``deviation of variable $y$ from its lower bound $y^L$, if $x
 \in [x_{k-1},x_k]$'' and is set to 0 if the value of $x$ lies outside of
 this range  \citep{gou09}.
 Figure~\ref{fig:mccormick}(b) shows how the relaxation improves when we
 break the domain of $x$ into two ranges $x_l .. x_m$ and $x_m .. x_u$.

\begin{flalign}
& \sum_{k = 1}^{K} \lambda_k  = 1 & & \label{eqn:SOS1}\\
& \sum_{k=1}^{K} x_{k-1} \, \lambda_k  \leq x \leq \sum_{k = 1}^{K} x_k \, \lambda_k \\
& y = y^L + \sum_{k = 1}^{K} \delta y_k \\
& 0 \leq \delta y_k \leq (y^U - y^L) \lambda_k & & \forall k \\
& z \geq y^U \, x + \sum_{k = 1}^{K} x_k \, \delta y_k - (y^U - y^L) \sum_{k = 1}^{K} x_k  \, \lambda_k \\
& x \leq y^U \, x + \sum_{k=1}^{K} x_{k-1} \, \delta y_k - (y^U - y^L) \sum_{k = 1}^{K} x_{k-1} \, \lambda_k \\
& z \leq y^L \, x + \sum_{k = 1}^{K} x_k \, \delta y_k\\
& z \geq y^L \, x + \sum_{k = 1}^{K}
\end{flalign}

In the \textsc{DistanceTo}$_{STV}$ MINLP, variables $\tau_j$ and $y_{\pi[j],j,s}$ (Constraint \eqref{eqn:define-d}), and $\tau_j$ and $\rho_j$ (Constraint \eqref{eqn:define-tv2}) participate in bilinear terms. We treat the transfer value of votes distributed from the candidate elected in round $j$, $\tau_j$, as $x$ in the piecewise linear relaxation scheme presented above, with variables $y_{\pi[j],j,s}$ (the total value of votes of signature $s$ in the tally of candidate $\pi[j]$ at the start of round $j$) and $\rho_j$ (the number of transferable votes in the tally of the candidate elected in round $j$) as $y$. In Section \ref{sec:Evaluation}, we discretise the domain of $\tau_j$ into varying numbers of segments (by varying $K$), and compare the resulting lower bounds (on margins) found by {\bf margin-stv} to the true MOV found when no such relaxation is used. 

In their algorithm for computing margins in IRV elections, \cite{Magrino:IRV} recognise that their \textsc{DistanceTo} LP, defined to compute the smallest number of vote changes required to realise a particular (partial or complete) elimination order, need only consider a subset of possible vote signatures (in place of all possible partial and total orderings over the set of candidates). Given an elimination order $\pi$, specifying which candidates are eliminated in each counting round, \cite{Magrino:IRV} note that classes of ballot signatures, denoted \textit{equivalence} classes, will behave in the same way (be transferred between candidates in the same way) as counting progresses.  Consequently, each \textsc{DistanceTo} LP need only contain variables defining the number of votes of each \textit{equivalence} class present in the modified election profile. The definition of equivalence classes presented by  \cite{Magrino:IRV} can be applied in the context of both IRV and STV elections, given an elimination sequence (for an IRV election) and a sequence of elections and eliminations (for an STV election). In our implementation of \textbf{margin-stv}, we define the \textsc{DistanceTo}$_{STV}$ MINLP (and its relaxations) over signature \textit{equivalence} classes (i.e., the set $\mathbf{S}$ contains the signatures of all \textit{equivalence} classes, in place of all partial and total orders over the set of candidates $\mathcal{C}$).     

\section{Evaluation} \label{sec:Evaluation}


\ignore{
\begin{table}[!t]
\centering
\begin{tabular}{r|r|r|crr}
\hline
\textbf{\#} & $|\mathcal{C}|$ & \textbf{Best Upper} & \textbf{MOV} & \textbf{MINLPs} & \textbf{Time (s)} \\
& & \multicolumn{1}{c|}{\textbf{Bound}} & & \textbf{solved} & \\
\hline
\hline
1 & 4 & 38,199 & 30,654 -- 37908 & 54 & 2,528 \\
2 & 4 & 5,683 & 2,725 -- 2,726 & 17 & 900 \\
3 & 4 & 6,530 & 6,359 -- 6,360 & 25 & 464 \\
4 & 4 & 2,527 & 63 & 18 & 171\\
5 & 4 & 653 & 59 & 17 & 20 \\
6 & 4 & 763 & 375 & 23 & 708 \\  
7 & 5 & 7,500 & 7,499 -- 7,500 & 48 & 213 \\
8 & 5 & 1,079 & 628 -- 629 & 45 & 2,972 \\
9 & 5 & 1,425 & 659 -- 660 & 38 & 2,331 \\
10 & 5 & 422 & 164 -- 165 & 42 & 565 \\
11 & 7 & 1,719 & 441 & 58 & 1 \\
12 & 7 & 2,289 & 1,326-- 1,327 & 2,247 & 13,693 \\
13 & 8 & 2,857 & 1,216 -- 1,576 & --- & $\infty$ \\
14 & 11 & 1,178 & 309 -- 1,178 & --- & $\infty$ \\
15 & 11 & 63 & 4 & 244 & 107 \\
16 & 13 & 2,188 & 823 -- 2,188 & --- & $\infty$ \\   
\hline
\end{tabular}
\caption{Application of {\bf margin-stv} to 16 US IRV elections re-imagined as STV elections with 2 seats,
  reporting: the number of candidates ($|\mathcal{C}|$); best upper bound on
  the MOV; computed MOV (or bounds on the MOV); number of MINLPs solved; and
  algorithm runtime (in seconds) ($\infty$ represents a time out of 12 hours).}
\label{tab:IRVasSTVexact}
\end{table}  }

We first evaluate {\bf margin-stv} on a set of small STV elections. To do so, we take 16 IRV elections conducted in the US between 2007 and 2010, and re-imagine them as STV elections with 2 available seats (Appendix \ref{sec:IRVlist} lists the names of each of these elections, alongside the number of candidates, and number of votes cast). The number of candidates in these elections range between 4 and 13. All experiments in this paper have been conducted on a machine with an Intel Xeon ES-2440 2.40GHz 6 core processor, and 64GB of RAM. 

Table \ref{tab:IRVasSTV} reports the results of {\bf margin-stv} -- \textit{without} and \textit{with} the use of the lower bounding rule of Section \ref{sec:SimpleLowerBound} -- on 16 small IRV (re-imagined as STV) elections. We record: the number of candidates $|\mathcal{C}|$ in the election; the best computed upper bound on the MOV (the minimum of the winner elimination and Simple-STV upper bound); the MOV returned by {\bf margin-stv} (or bounds on the MOV); the number of \textsc{DistanceTo}$_{STV}$ MINLPs solved by the algorithm; and the time taken by {\bf margin-stv} to compute the MOV (in seconds). SCIP \citep{Achterberg2009} is used to solve all \textsc{DistanceTo}$_{STV}$ MINLPs.  We terminate each run of {\bf margin-stv} after 12 hours, and infer a lower and upper bound on the MOV from the state of the frontier. Our algorithm did not terminate for three instances (13, 14, and 16) within 12 hours. In 11/16 instances, our algorithm could not produce an exact MOV, but a lower and upper bound on the MOV. In 7/11 instances, the lower and upper bound differed by only 1 vote (i.e., {\bf margin-stv} was able to produce very tight bounds on the MOV). Table \ref{tab:IRVasSTV} shows that our lower bounding rule can, in some instances, reduce the runtime of the algorithm. For instances 1, 2, 10, and 12, runtimes are reduced by 107, 761, 106, and 1,635 seconds. Our lower bounding rule can increase the runtime of the algorithm---for instances 3 and 6, runtimes are increased by 185 and 378 seconds. Our lower bounding rule consistently reduces the number of MINLPs solved, but if these MINLPs are quickly deemed infeasible by SCIP \citep{Achterberg2009}, the additional time spent computing lower bounds for each partial order visited becomes an overhead.

\begin{table}[!t]
\centering
\begin{tabular}{r|r|r|crr|crr}
  \hline
  &&& \multicolumn{3}{c|}{\textbf{margin-stv}} &
  \multicolumn{3}{c}{\textbf{margin-stv} with LB of
    Sec~\ref{sec:SimpleLowerBound}} \\
\textbf{\#} & $|\mathcal{C}|$ & \textbf{Best UB} & \textbf{MOV} & \textbf{MINLPs} & \textbf{Time}& \textbf{MOV} & \textbf{MINLPs} & \textbf{Time} \\
& & & & \textbf{solved} & \textbf{(s)} & & \textbf{solved} & \textbf{(s)} \\
\hline
\hline
1 & 4 & 38,199 & \bf 30,654 -- 37,908 & 54 & 2,528      & \bf 30,654 -- 37,908 & 35 & 2,421 \\        
2 & 4 & 5,683 & \bf 2,725 -- 2,726 & 17 & 900          & \bf 2,725 -- 2,726 & 7 & 139 \\             
3 & 4 & 6,530 & \bf 6,359 -- 6,360 & 25 & 464          & \bf 6,359 -- 6,360 & 19 & 649 \\            
4 & 4 & 2,527 & \bf 63 & 18 & 171                      & \bf 63 & 14 & 169 \\                        
5 & 4 & 653 & \bf 59 & 17 & 20                         & \bf 59 & 11 & 19 \\                         
6 & 4 & 763 & \bf 375 & 23 & 708                       & \bf 375 & 13 & 1,086 \\                     
7 & 5 & 7,500 & \bf 7,499 -- 7,500 & 48 & 213          & \bf 7,499 -- 7,500 & 30 & 211 \\            
8 & 5 & 1,079 & \bf 628 -- 629 & 45 & 2,972            & \bf 628 -- 629&  30 & 2,972 \\              
9 & 5 & 1,425 & \bf 659 -- 660 & 38 & 2,331            & \bf 659 -- 660 & 26 & 2,318 \\              
10 & 5 & 422 & \bf 164 -- 165 & 42 & 565               & \bf 164 -- 165 & 22 & 459 \\                
11 & 7 & 1,719 & \bf 441 & 58 & 1                      & \bf 441 & 39 & 1 \\                         
12 & 7 & 2,289 & \bf 1,326-- 1,327 & 2,247 & 13,693    & \bf 1,326 -- 1,327 &  1,006 &  12,058 \\    
13 & 8 & 2,857 & 1,216 -- \bf 1,576 & --- & $\infty$   & \bf 1,309 -- 1,576 & --- & $\infty$ \\      
14 & 11 & 1,178 & 309 -- \bf 1,178 & --- & $\infty$    &  \bf 351 -- 1,178 & --- & $\infty$ \\       
15 & 11 & 63 & \bf 4 & 244 & 107                       & \bf 4 & 190 & 106 \\                        
16 & 13 & 2,188 & \bf 823 -- 2,188 & --- & $\infty$    & 0 -- \bf 2,188 & --- & $\infty$ \\          
\hline
\end{tabular}
\caption{Application of {\bf margin-stv} to 16 US IRV elections re-imagined as STV elections with 2 seats,
  reporting: the number of candidates ($|\mathcal{C}|$); best upper bound on
  the MOV; computed MOV (or bounds on the MOV); number of MINLPs solved; and
  algorithm runtime (in seconds) ($\infty$ represents a time out of 12
  hours) without and with  the use of the lower bounding rule of
  Sec. \ref{sec:SimpleLowerBound}.
  Best results for bounds are bolded.}
\label{tab:IRVasSTV}
\end{table}

\begin{table}[!tb]
\centering
\begin{tabular}{r|r|c|rrr|rrr}
\hline
  &&& \multicolumn{3}{c|}{\textbf{margin-stv}} &
  \multicolumn{3}{c}{\textbf{margin-stv} with LB of
    Sec~\ref{sec:SimpleLowerBound}} \\
\textbf{\#} & $|\mathcal{C}|$ & \textbf{Exact MOV} & \textbf{MOV} & \textbf{MIPs} & \textbf{Time}& \textbf{MOV} & \textbf{MIPs} & \textbf{Time} \\
& & \textbf{Bounds} &\textbf{(LB)} & \textbf{solved} & \textbf{(s)} &\textbf{(LB)} & \textbf{solved} & \textbf{(s)}\\
\hline
\hline
1 & 4 & 30,654 -- 37,908 & 17,662 & 28 & 0.8   & \bf 23,079 & 11 & 0.2 \\         
2 & 4 & 2,725 -- 2,726  & 399 &  15& 0.2       & \bf 789 &  5& 0.1\\           
3 & 4 & 6,359 -- 6,360 & 3245 & 28 & 0.4       & \bf 6356 & 12 & 0.1\\         
4 & 4 & 63  & \bf 0 &  12 & 0.1                    & \bf 0 &  8 & 0.1\\            
5 & 4 & 59 & \bf 0& 13 & 0.1                       & \bf 0& 7 & 0.1\\              
6 & 4 & 375 & 43 & 19 & 0.1                    & \bf 46 & 10 & 0.1\\           
7 & 5 & 7,499 -- 7,500 & \bf 0& 23 & 0.3           & \bf 0& 13 & 0.2\\             
8 & 5 & 628 -- 629 & 218& 33 & 0.3             & \bf 339& 20 & 0.3\\           
9 & 5 & 659 -- 660 & \bf 64 & 26 & 0.3             & \bf 64 & 17 & 0.3\\           
10 & 5 & 164 -- 165 & 34& 25 & 0.2             & \bf 91& 19 & 0.4\\            
11 & 7 & 441 & \bf 441 & 57 & 0.6                  & \bf 441 & 38 & 0.6\\          
12 & 7 & 1,326 -- 1,327 & 413 & 48 & 0.9       & \bf 859 & 166 & 4 \\          
13 & 8 & 1,309 -- 1,576  & 289 &  215 &  8     & \bf 289 &  117 &  4\\         
14 & 11 & 351 -- 1,178  & \bf 0 &  723& 217        & \bf 0 &  347& 36\\            
15 & 11 & 4 & \bf 4 & 243& 8                       & \bf 4 & 189& 11\\             
16 & 13 & 823 -- 2,188 & \bf 0 & --- & $\infty$    & \bf 0 & 21,520 & 23,243\\     
\hline
\end{tabular}
\caption{Application of relaxed {\bf margin-stv}  (bilinear terms replaced
  with \cite{mcc76} inequalities, \textit{without}
  and \textit{with} the lower bounding rule of Sec.
  \ref{sec:SimpleLowerBound})
  to the STV elections of Table \ref{tab:IRVasSTV}, reporting: the
  number of candidates ($|\mathcal{C}|$); the
  best known bounds on MOV from Table \ref{tab:IRVasSTV}; lower bound on the MOV
  computed by {\bf margin-stv}; number of MIPs solved; and
  algorithm runtime (in seconds). Best lower bounds are bolded.}
\label{tab:McCormicksBoth}
\end{table}

Table \ref{tab:McCormicksBoth} reports the results of the relaxed \textbf{margin-stv} algorithm (with bilinear terms replaced by \cite{mcc76} inequalities) in each STV instance of Table \ref{tab:IRVasSTV}, \textit{without} and \textit{with} the lower bounding rule of Section \ref{sec:SimpleLowerBound}. CPLEX 12.5 is used to solve all \textsc{DistanceTo}$_{STV}$ MIPs formed by our {\bf margin-stv} relaxations. We report the lower bound on the MOV found by \textbf{margin-stv} alongside the exact MOV (or bounds on the exact MOV) reported in Table \ref{tab:IRVasSTV}. The lower bounds found by \textbf{margin-stv} (with \cite{mcc76} inequalities) are often significantly lower than the exact MOV. Our lower bounding rule is beneficial in this setting -- it often discovers tighter lower bounds on partial and complete orders than the MIP relaxation of \textsc{DistanceTo}$_{STV}$ (in 7/16 instances, applying the lower bounding rule results in a tighter lower bound being found by relaxed \textbf{margin-stv}), and reduces runtimes significantly for larger instances. 

\begin{table}[!tb]
\centering
\small
\begin{tabular}{r|c|rrr|rrr}
\hline
& & \multicolumn{3}{c|}{$K = 5$} & \multicolumn{3}{c}{$K = 10$}\\
\cline{3-8}
\textbf{\#} & \textbf{Exact MOV} & \textbf{MOV} & \textbf{MIPs} & \textbf{Time} & \textbf{MOV} & \textbf{MIPs} & \textbf{Time} \\
& \textbf{Bounds} & \textbf{LB}& \textbf{solved} & \textbf{(s)} & \textbf{LB}& \textbf{solved} & \textbf{(s)}  \\
\hline
\hline
 1 & 30,654 -- 37,908 & 28,038 & 12 & 1 & 29,020& 12 & 1\\
 2 & 2,725 -- 2,726 & 2,265 & 6 & 0.5 & 2,512 & 6 & 1\\
 3 & 6,359 -- 6,360 & 6,356 & 13 & 0.4 & 6,356 & 17 & 2 \\
 4 &63 & \bf 0 & 8 & 0.1 & \bf 0 & 8 & 0.1\\
 5 &59 & 0 & 7 & 0.4 & 10 & 9& 1\\
 6 &375 & 327 & 10 & 0.2 & 337 & 10 & 0.6\\
 7 &7,499 -- 7,500 & 5,168 & 14 & 0.9 & 6,347 & 28 & 4\\
 8 &628 -- 629 & 531 & 24 & 2 &581  & 25& 7\\
 9 &659 -- 660 & 558 & 19 & 0.9 & 609 & 19& 3 \\
 10 &164 -- 165 & 133 & 19 & 4& 141 & 19 & 3\\
 11 &441 & \bf 441 & 38 &  0.7& \bf 441& 38& 0.8\\
 12 &1,326 -- 1,327 & 1049 & 486 & 37 & 1,211 & 698& 83\\
 13 &1,309 -- 1,576 & 1321 & 538 & 65 & 1,431 & 636 & 243\\
 14 &351 -- 1,178 & 53 & 84 & 22& 602 & 3,008& 2,900\\
 15 & 4& \bf 4 & 189 & 22& \bf 4 & 189 & 27\\
 16 &823 -- 2,188 & \bf 760 & 7,025& 35,583 & 95 & ---  & $\infty$ \\ 
\hline
& & \multicolumn{3}{c|}{$K = 15$} & \multicolumn{3}{c}{$K = 20$} \\
\cline{3-8}
\textbf{\#} & \textbf{Exact MOV} & \textbf{MOV} & \textbf{MIPs} & \textbf{Time} & \textbf{MOV} & \textbf{MIPs} & \textbf{Time} \\
& \textbf{Bounds} & \textbf{LB}& \textbf{solved} & \textbf{(s)} & \textbf{LB}& \textbf{solved} & \textbf{(s)} \\
\hline
\hline
 1 & 20,654 -- 37,908 & 29,549 & 12 & 3 & \bf 29,968 & 12& 3\\
 2 &2,725 -- 2,726 & 2,582 & 6 & 2 & \bf 2,609 & 6 & 1\\
 3 & 6,359 -- 6,360 & 6,356 & 17 & 2 & \bf 6,356 & 17 & 1\\
 4 & 63& 0 & 11 & 0.4& \bf 0 & 11 & 1\\
 5 & 59& 28 & 9 & 1 & \bf 35 & 9& 1\\
 6 & 375& 351 & 10& 1& \bf 358 & 10 & 2 \\
 7 & 7,499 -- 7,500& 6,721 & 29 & 3& \bf 6,923 & 29& 4 \\
 8 &628 -- 629 & 598& 25& 10& \bf 606 & 25& 12\\
 9 &659 -- 660 & 623 & 19& 5& \bf 638& 20& 4 \\
 10 &164 -- 165 & 147 & 19& 6& \bf 155 & 19& 6\\
 11 &441  & \bf 441 & 38& 0.6& \bf 441 & 38 & 1\\
 12 &1,326 -- 1,327 & 1,221& 700& 105& \bf 1,257 & 758 & 185\\
 13 &1,309 -- 1,576 & 1,468& 747& 452& \bf 1,486& 754 & 638\\
 14 &351 -- 1,178 & 785& 10,511& 15,732& \bf 878& 13,786& 28,909\\
 15 &4 & \bf 4 & 189 & 33& \bf 4 & 189 & 39\\
 16 &823 -- 2,188 & 17 & --- & $\infty$ & 0& --- & $\infty$\\
\hline
\end{tabular}
\caption{Application of piecewise-relaxed {\bf margin-stv}  (\textit{with}
  the lower bounding rule of Sec. \ref{sec:SimpleLowerBound}) to
  the STV elections of Table \ref{tab:IRVasSTV}, reporting: the best
  known bounds MOV from Table \ref{tab:IRVasSTV};
  lower bound on the MOV computed by  piecewise-relaxed {\bf margin-stv};
   number of MIPs solved; and algorithm runtime (in seconds).}
\label{tab:PwiseWBounds}
\end{table}

Relaxed \textbf{margin-stv}  (with \cite{mcc76} inequalities and our lower bounding rule) can compute a lower bound on the margin in each of our 16 STV instances within 12 hours. However, in 5/16 instances a trivial lower bound of 0 is found. We consider whether solving a piecewise linear relaxation of each \textsc{DistanceTo}$_{STV}$ MINLP, as described in Section \ref{sec:ImplementationDetails}, results in  \textbf{margin-stv} finding better lower bounds.  Table \ref{tab:PwiseWBounds} reports the results of piecewise-relaxed \textbf{margin-stv} (with the use of our lower bounding rule) given varying $K$ (with higher values of $K$ forming a more accurate relaxation of each \textsc{DistanceTo}$_{STV}$ MINLP). In general, the resulting MOV lower bounds increase as $K$ increases. The counter-example is instance 16, for which \textbf{margin-stv} computes lower bounds of 760, 95, 17, and 0 for $K$ $=$ 5, 10, 15, and 20. As $K$ increases, the MIP relaxations solved throughout the algorithm become more complex and time consuming to solve. Consequently, \textbf{margin-stv} makes less progress through the space of partial orders in instance 16 within the 12 hour time limit. In the $K = 5$ setting, \textbf{margin-stv} finds reasonable lower bounds in most instances. The exact MOV is found in 2 instances (11 and 15). In instance 13, we find a margin that lies within the bounds found on the MOV by the exact algorithm. Across instances 1 to 3, 6, 8 to 10, 12, and 16, the MOV lower bounds discovered are, on average, 14.4\% below the exact MOV (or lower bound on the MOV found by the exact algorithm). In instances 4 and 5, piecewise-relaxed \textbf{margin-stv} finds trivial lower bounds of 0, however the exact MOV in these instances is  small (63 and 59 votes, respectively). 

\subsection{Real-World STV Elections}\label{sec:RWSE}

In the preceding section we have evaluated \textbf{margin-stv} on a series of IRV elections (re-imagined as STV elections with 2 seats) held in the US between 2007 and 2010. The number of candidates in these elections range from 4 to 13. STV elections often involve a large number of candidates. In the 2012 election of senators to the ACT senate in Australia, 5-7 senators were elected in each of 3 districts. An STV election was held for each district, involving 20, 26, and 28 candidates. 
In the 2014 Victorian senate election in Australia, 5 senators were elected in each of 5 districts. The number of candidates in these elections ranged from 37 to 52. The use of \textbf{margin-stv} for elections of this size, without relaxation, is not feasible (due to the complexity of the \textsc{DistanceTo}$_{STV}$ MINLPs that must be solved). We demonstrate the value of our \textbf{margin-stv} algorithm for computing \textit{lower bounds} on the  margin in a number of real-world STV elections.   

In addition to the use of MINLP relaxations (\cite{mcc76} inequalities and piecewise-linear relaxations), we parallelise the exploration of the frontier of partial candidate orders. In place of selecting \textit{one} partial candidate order to expand (at Step 13 of Figure \ref{alg:MarginSTV}), we select the first $N_F$ (where $N_F$ is a parameter) orders in the frontier $F$. Each of these orders is expanded in parallel, and the frontier updated accordingly -- the children of \textit{all} expanded orders with \textsc{DistanceTo}$_{STV}$ evaluations that are less than the current upper bound are inserted into the frontier. This upper bound is revised, and orders pruned from the frontier, when a complete order is found with a  \textsc{DistanceTo}$_{STV}$ evaluation that is lower than the current upper bound. 

The number of variables and constraints in the \textsc{DistanceTo}$_{STV}$ models solved by \textbf{margin-stv} is determined  by the number of candidates in the orders being evaluated. In most STV elections, the number of available seats is much less than the number of candidates. Consequently, these orders will contain lengthy sequences of candidate eliminations. To limit the memory and solve time requirements of the \textsc{DistanceTo}$_{STV}$ models formed by \textbf{margin-stv}, we group these sequences of eliminated candidates together, ignoring constraints that enforce their relative elimination order (i.e., these candidates are effectively eliminated in a single round of counting). \ignore{This allows us to take a 28 candidate order, for example, and reduce its complexity to that of a 8 candidate order (with 8 rounds of candidate election and elimination).} Grouping eliminated candidates together reduces the number of constraints required, and significantly reduces the number of vote \textit{equivalence} classes that need to be considered, in each \textsc{DistanceTo}$_{STV}$ model. This grouping adds an additional level of relaxation to the algorithm.

We consider, in this section, the performance of {\bf margin-stv} on 28 STV elections. We consider: the 2013 and 2016 election of 2 candidates to the Federal Senate for the Australian Capital Territory (ACT) and Northern Territory (NT); the 2002 election of 4, 3, and 5 candidates to the lower house of the Irish parliament (the D\'{a}il \'{E}ireann) representing the Dublin North, Dublin West, and Meath constituencies; and 21 STV elections electing 3 to 4 candidates (out of 8 to 13, per ward) to the Glasgow City Council in 2007. Data for the Irish and Scottish STV elections were obtained from PrefLib\footnote{www.preflib.org}. In each experiment reported in this section, {\bf margin-stv} has been afforded a time limit of 24 hours. If this time limit is reached, and the algorithm has not converged, we report the best lower bounds on the MOV computed in that time period. Unless otherwise stated, we set $N_F = 5$ and run {\bf margin-stv} with increasing values of $K$ (from $K = 5$ to $K = 100$) until the resulting lower bound does not improve, or is within 10\% of the best known upper bound.

\subsection{2002 Irish General Election}

We first consider the 2002 election of 4, 3, and 5 candidates (out of 13, 9, and 15) to the Irish lower house to represent Dublin North, Dublin West, and Meath. Table \ref{tab:IRISH} reports, for each election, the quota, number of votes cast, and the best known upper bound on the MOV (computed as per Figure \ref{alg:WEUB} and Section \ref{sec:SimpleBounds}). For increasing values of $K$, we report the lower bound on the MOV computed by {\bf margin-stv} over a 24 hour period, the runtime (in seconds) of the algorithm (if less than 24 hours), and the number of MIPs solved.

\begin{table}[t]
\centering
\begin{tabular}{l|c|c|c|c|c|c|c|c|c}
\hline
{\bf Election} & {\bf Seats} & $|\mathcal{C}|$ & {\bf Quota} & {\bf Votes} & {\bf Upper}  & $K$ & {\bf Lower}   & {\bf Time (s)} & {\bf MIPs}\\
         &       &            &       &  {\bf cast} & {\bf Bound}     &     &   {\bf Bound}     &        & {\bf Solved}\\
         &       &            &       &    & {\bf on MOV}     &     &   {\bf on MOV}     &        & \\
\hline
\hline
Dublin North &  4 & 13    & 8,789 & 43,942 &  211 & 5 &  0  & 46 & 400 \\
             &    &       &       &        &      & 15 & 181 & 2,905 & 487 \\    
             &    &       &       &        &      & 25 & 185 & 10,808 & 276  \\  
             &    &       &       &        &      & 35 & 199 & 50,994 & 394   \\
\hline
Dublin West &  3  &  9  &  7,498 & 29,988 & 385   & 5 &  0   & 15  & 143 \\
            &     &     &        &        &       & 15&  153 & 241 & 147 \\
            &     &     &        &        &       & 25&  211 & 506 &  153  \\
            &     &     &        &        &       & 35&  235 & 1,628&  151  \\
            &     &     &        &        &       & 60&  257 & 3,914 & 155   \\
            &     &     &        &        &       & 80&  267 & 14,649 & 161  \\
            &     &     &        &        &       & 100& 269 & 38,660&  157  \\
\hline
Meath & 5 & 15 & 10,681 & 64,081 & 1,135 & 5 & 275 & -- & 1,997 \\
      &   &    &        &        &       & 15&  275 & 72,349   &  549 \\    
\hline
\end{tabular}
\caption{The winner elimination upper bound, and lower bound on the MOV, computed for 3 STV elections held in the 2002 Irish General Election (for varying values of $K$). The number of seats, candidates, quota, and number of votes cast in each election is reported, alongside the runtime (in seconds) of {\bf margin-stv} (-- indicates that the algorithm reached the 24 hour timeout) and the number of MIPs solved.} 
\label{tab:IRISH}
\end{table}

{\footnotesize
\begin{table}[!htbp]
\centering
\begin{tabular}{l|c|c|c|c|c|c|c|c|c}
\hline
{\bf Election} & {\bf Seats} & $|\mathcal{C}|$ & {\bf Quota} & {\bf Votes} & {\bf MOV}  & $K$ & {\bf MOV}   & {\bf Time } & {\bf MIPs}\\
         &       &            &       &  {\bf cast} & {\bf Upper}     &     &   {\bf Lower}     &    {\bf (s)}    & {\bf Solved}\\
         &       &            &       &    & {\bf Bound}     &     &   {\bf Bound}     &        & \\
\hline
\hline
Linn &  4 & 11 & 1,914 & 9,567  &   221    &   5  & 94 &  1,147  & 1,840 \\
          &   &    &       &       &     & 15  & 139 & --   &  5,867   \\
\hline 
Newlands & 3  & 9 & 2,164  &  8,654  &  88   & 5  & 47 &  14  & 203 \\
          &   &    &       &       &     & 15  & 76 &  78  &  284   \\
          &   &    &       &       &     & 25  & 81 &  363  &   284  \\
\hline
Greater Pollock & 4 & 9 & 1,737   & 8,682  &  223   & 5  & 125 &  1,866  & 1,647 \\
          &   &    &       &       &     & 15  & 197 & 38,428   &  3,349   \\
          &   &    &       &       &     & 25  & 159* & --   &  2,304   \\
\hline
Craigton & 4 & 10 & 2,211 & 11,052 & 93 & 5 & 0 & 62 & 174 \\
          &   &    &       &       &     & 15  & 0 &  493  &   175  \\
          &   &    &       &       &     & 25  & 25 &  1,176  &  168   \\
          &   &    &       &       &     & 35  & 37 &  11,321  &  179   \\
          &   &    &       &       &     & 50  & 43 &  39,892  &  188   \\
          &   &    &       &       &     & 80  & 31* &  --  &  164   \\
\hline
Govan & 4 & 11 & 1,913&  9,560   &   278  & 5  &  149 &  --  & 10,808 \\
          &   &    &       &       &     & 15  & 131* &  --  & 5,226    \\
\hline
Pollockshields & 3 &  9 & 2,392  & 9,567 &  30   & 5  & 0 &   1 & 39 \\
          &   &    &       &       &     & 15  & 0 &  6  &  39   \\
          &   &    &       &       &     & 25  & 2 &  19  &  39   \\
          &   &    &       &       &     & 35  & 4 &  20  &  39   \\
          &   &    &       &       &     & 80  & 4 &  347  &  39   \\
          &   &    &       &       &     & 100  & 6 &  681  &  39   \\
\hline
Langside &  3 & 8 & 2,334  &  9,334  &   246  & 5  &  124&  25  &  224\\
          &   &    &       &       &     & 15  & 193 &  555  &  545   \\
          &   &    &       &       &     & 25  & 209 &  3,350  &  553   \\
          &   &    &       &       &     & 35  & 217 &  7,266  & 742    \\
          &   &    &       &       &     & 45  & 221 &  14,495  & 746    \\
          &   &    &       &       &     & 55  & 224 &   19,341 &  752   \\
\hline
Southside Central & 4 & 9  & 1,748   &   8,738    &   231  & 5  & 113 & 1,306 & 1,418 \\
          &   &    &       &       &     & 15  & 187 &  68,537  &  2,313   \\
          &   &    &       &       &     & 25  & 118* &  --  &  1,120   \\
\hline
Calton &  3 & 10 & 1,300 & 5,199 & 959 & 5 & 215 & 18,369 & 39,356 \\
          &   &    &       &       &     & 15  & 130* &  --  &  20,048   \\
\hline
Anderston & 4 &  9 & 1,381 & 6,900 & 106 & 5 & 3 & 57 & 168 \\
          &   &    &       &       &     & 15& 64 & 5,596 & 346 \\
          &   &    &       &       &     & 25& 85 & 16,784   &  407   \\
          &   &    &       &       &     & 35 & 94 &  51,571  &  517   \\
\hline
Hillhead & 4 & 10 &  1,797  &  8,984 &  112   & 5  & 0 &   35 & 271 \\
          &   &    &       &       &     & 15  & 41 &  1,621  &  270   \\
          &   &    &       &       &     & 25  & 55 &  8,277  &  270   \\
          &   &    &       &       &     & 35  & 58 &  27,949  &  265   \\
          &   &    &       &       &     & 45  & 0* &  --  &  251   \\
\hline
\end{tabular}
\caption{The best known upper bound, and lower bound on the MOV, computed for the first 11/21 STV elections held in the 2007 Glasgow City Council Election (for varying values of $K$). The number of seats, candidates, quota, and number of votes cast in each election is reported, alongside the runtime (in seconds) of {\bf margin-stv} (-- indicates that the algorithm reached the 24 hour timeout) and the number of MIPs solved.} 
\label{tab:GLASGOW1}
\end{table}}

{\footnotesize
\begin{table}[!htbp]
\centering
\begin{tabular}{l|c|c|c|c|c|c|c|c|c}
\hline
{\bf Election} & {\bf Seats} & $|\mathcal{C}|$ & {\bf Quota} & {\bf Votes} & {\bf MOV}  & $K$ & {\bf MOV}   & {\bf Time} & {\bf MIPs}\\
         &       &            &       &  {\bf cast} & {\bf Upper}     &     &   {\bf Lower}     &    {\bf (s)}    & {\bf Solved}\\
         &       &            &       &    & {\bf Bound}     &     &   {\bf Bound}     &        & \\
\hline
\hline
Partick West & 4 & 9  & 2,549   &  12,744   &   194  & 5  & 18 &  22  & 166 \\
          &   &    &       &       &     & 15  &  133 &  728  &  207   \\
          &   &    &       &       &     & 25  & 174 &  6,407  &  311   \\
          &   &    &       &       &     & 35  & 187 &  10,343  & 354    \\
\hline
Garscadden & 4 & 10 &  2,033&  10,160 & 449  &   5  & 197 &  --  &  5,080 \\
          &   &    &       &       &     & 15  & 96* &  --  &   780  \\
\hline
Drumchapel & 4 & 10 & 1,737 & 8,680 & 1,325 & 5 & 335 & 19,325 & 16,257 \\
          &   &    &       &       &     & 15  &  0*&  --  &  353   \\
\hline
Maryhill & 4 & 8 & 1,981  &  9,901 &  292   & 5  & 70 &  140  & 329 \\
          &   &    &       &       &     & 15  & 203 &  11,580  &  606   \\
          &   &    &       &       &     & 25  & 226 &  25,849  &  776   \\
          &   &    &       &       &     & 35  & 234 &  78,380  &  747   \\
	 &   &    &       &       &     & 45  & 194* & --   & 482    \\
\hline
Canal & 4 & 11 & 1,725 & 8,624 & 148 & 5 & 68 & 487 & 576 \\
          &   &    &       &       &     & 15  & 98 &  9,384  &  729   \\
          &   &    &       &       &     & 25  & 108 &  36,690  &  852   \\
          &   &    &       &       &     & 35  & 109 &  79,946  &  844   \\
          &   &    &       &       &     & 45  & 69* &  --  &  300   \\
\hline
Springburn & 3 & 10  & 1,353  & 5,410  &  1,014   & 5  & 447  &  22,772  & 37,236 \\
          &   &    &       &       &     & 15  & 305* &  --  & 3,979    \\
\hline 
East Centre & 4 & 13 & 1,816 & 9,078 & 134 & 5 & 73 & 26,753 & 16,360 \\
          &   &    &       &       &     & 15  & 71* & --   &  10,389   \\
\hline
Shettleston & 4 & 11 &  1,761  &  8,803  &  355   & 5  &  158 &  --  &  13,014\\
          &   &    &       &       &     & 15  & 145* &  --  &  4,817   \\
\hline
Baillieston & 4 &  11 &  2,076 &  10,376  & 108 &  5  & 0 & 457   & 266 \\
          &   &    &       &       &     & 15  & 38 &  2,088  &  347   \\
          &   &    &       &       &     & 25  &46  & 11,122   &  345   \\
          &   &    &       &       &     & 35  & 58 &  49,942  &  383   \\
          &   &    &       &       &     & 45  & 64 &  74,870  &  390   \\
          &   &    &       &       &     & 55  & 8* &  --  &  300   \\
\hline
North East & 4 & 10 &  1,673&   8,363    &  911   & 5  &  275 &  52,653  &  9,967\\
          &   &    &       &       &     & 15  & 60* & --   &  385   \\
\hline
\end{tabular}
\caption{The best known upper bound, and lower bound on the MOV, computed for the second 10/21 STV elections held in the 2007 Glasgow City Council Election (for varying values of $K$). The number of seats, candidates, quota, and number of votes cast in each election is reported, alongside the runtime (in seconds) of {\bf margin-stv} (-- indicates that the algorithm reached the 24 hour timeout) and the number of MIPs solved.} 
\label{tab:GLASGOW2}
\end{table}}

\subsection{2007 Glasgow City Council}

In the 2007 Glasgow City Council election, 3 to 4 candidates (out of 8 to 13) were elected in 21 STV elections (one for each of 21 wards). Tables \ref{tab:GLASGOW1} and \ref{tab:GLASGOW2} report the quota, number of votes cast, and the best known upper bound on the MOV (computed as per Figure \ref{alg:WEUB} and Section \ref{sec:SimpleBounds}). For increasing values of $K$, we report the lower bound on the MOV computed by {\bf margin-stv} over a 24 hour period, the runtime (in seconds) of the algorithm (if less than 24 hours), and the number of MIPs solved. Instances where the MOV lower bound computed within 24 hours reduces as $K$ is increased are marked with an asterisk. Increasing $K$ increases the complexity, and solve times, of the MIPs being solved. These instances demonstrate that, as $K$ increases, the branch-and-bound search conducted by {\bf margin-stv} may make less progress over the 24 hour period, resulting in a lower (rather than improved) lower bound on the MOV.

\subsection{2013 and 2016 NT and ACT Federal Senate}

Table \ref{tab:AUS} reports lower bounds on the MOV computed for the 2013 and 2016 ACT and NT Federal Senate elections. While these elections have a small number of seats (2), there are many candidates (between 19 and 27). In each of these elections, the two winning candidates have orders of magnitude more votes in their initial tallies than all others. We expect the margin in these elections to be reasonably high (certainly greater than a few hundred votes) as we need to change enough votes to ensure one of these candidates is not awarded a seat (and that one alternate candidate \textit{is} awarded a seat). The MOV lower bounds found by {\bf margin-stv} in Table \ref{tab:AUS} are likely to be much lower than the true margins. 

\begin{table}[t]
\centering
\begin{tabular}{l|c|c|c|c|c|c|c|c|c}
\hline
{\bf Election} & {\bf Seats} & $|\mathcal{C}|$ & {\bf Quota} & {\bf Votes} & {\bf Upper}  & $K$ & {\bf Lower}   & {\bf Time (s)} & {\bf MIPs}\\
         &       &            &       &  {\bf cast} & {\bf Bound}     &     &   {\bf Bound}     &        & {\bf Solved}\\
         &       &            &       &    & {\bf on MOV}     &     &   {\bf on MOV}     &        & \\
\hline
\hline
2013 NT &  2 & 24 & 34,494 & 103,479 & 25,777 & 5 &  223  & -- & 561,999 \\
\hline
2016 NT &  2 & 19 & 34,010 & 102,027 & 23,373 & 5 &  3121  & -- & 219,876 \\
\hline
2013 ACT &  2 & 27 & 82,248 & 246,742 & 35,476& 5 &  82  & -- & 112,629 \\
\hline
2016 ACT &  2 & 22 & 84,923 & 254,767 & 44,499 & 5 &  205 & -- & 86,825 \\
\hline
\end{tabular}
\caption{The best known upper bound, and lower bound on the MOV, computed for the 2013 and 2016 NT and ACT Federal Senate elections  (for varying values of $K$). The number of seats, candidates, quota, and number of votes cast in each election is reported, alongside the runtime (in seconds) of {\bf margin-stv} (-- indicates that the algorithm reached the 24 hour timeout) and the number of MIPs solved.} 
\label{tab:AUS}
\end{table}

The elections of Table \ref{tab:AUS} follow a similar pattern -- a candidate is awarded a seat in the first round of counting, and then a long sequence of candidates are eliminated before a second candidate achieves a quota and is awarded a seat. In these circumstances, we can compute a lower bound on the MOV under the assumption that we cannot change who is elected and eliminated in the first $R$ rounds of counting. For large values of $R$, this assumption prunes significant portions of the space of possible elimination and election sequences from consideration. Consider the 2013 ACT election, in which the two available seats are awarded in rounds 1 and 24. If we assumed the first 20 rounds ($R = 20$) are fixed (i.e., the first winner is elected in round 1, and the candidates eliminated in rounds 2 to 20 are still eliminated, at some point, between rounds 2 to 20), Table \ref{tab:ACT13fixed} reports our computed lower bounds on the MOV for this election. 

\begin{table}[ht]
\centering
\begin{tabular}{l|c|c|c}
\hline
$K$ & {\bf Lower Bound } & {\bf Time (s)} & {\bf MIPs}\\
    &   {\bf on MOV}     &        & {\bf Solved}\\
\hline
\hline
 5 &  5879 &  117& 41 \\
 25 & 5879  & 4899 &  111\\
 50 & 5911  & 12,386 & 112 \\
80  & 5919  & 18,604 &  112\\
\hline
\end{tabular}
\caption{Lower bounds on the MOV, computed for the 2013 ACT Federal Senate election, for varying values of $K$, under the assumption that the first $R = 20$ rounds of the original count are fixed. For this election, there are: 2 seats; 27 candidates; a quota of 82,248 votes; and an upper bound on the MOV of 35,476 votes.} 
\label{tab:ACT13fixed}
\end{table}

The lower bounds reported in Table \ref{tab:ACT13fixed}, for the 2013 ACT Federal Senate election, are not guaranteed lower bounds on the true margin of victory of the election. If we take the minimal manipulation discovered by {\bf margin-stv} in the $K = 80$ case, and simulate the modified election with the suggested vote changes put in place, we find that the original winning candidates ($c^1_{w}$ and $c^2_w$) are given a seat (in rounds 1 and in the last round). However, the tallies of $c^2_w$ and the runner up $c_r$, at the point at which $c_r$ is eliminated differ in only 159 votes. We consider an additional manipulation that awards 80 additional first preference votes to the runner up $c_r$, and removes 80 first preference votes from $c^2_w$ (we effectively swap 80 votes that ranked $c^2_w$ first with a ranking that has $c_r$ in first place). Simulating this modified election gives us a different outcome, with candidates $c^1_{w}$ and  $c_r$ elected to a seat (in rounds 1 and in the last round). For this election, we now know that the MOV is no more than 5,999 votes -- we were able to find an alternate outcome by modifying 5,999 votes to an alternate ranking. The upper bounds found by the winner elimination upper bound method of Figure \ref{alg:WEUB}, and the simple upper bound method of Section \ref{sec:SimpleBounds}, are much higher than the true MOV. This analysis suggests there are alternate ways of using {\bf margin-stv} for computing lower and upper bounds on the MOV for large, and challenging, elections (with many seats or large numbers of candidates). In future work, we aim to explore this further.

\section{Concluding Remarks}

In this paper we develop an algorithm, \textbf{margin-stv}, for computing exact margins of
victory in STV elections, assuming the use of the Inclusive Gregory method
of surplus distribution. The algorithm is based on solving a mixed-integer
nonlinear formulation for the problem of computing a minimal manipulation to
achieve a desired election order.
We are able to compute exact margins of victory for small
elections, e.g. less than 12 candidates and 2 seats.
For larger, real elections \textbf{margin-stv}
is able to compute reasonable lower bounds within 24 hours (as demonstrated on 28 real-world STV elections).
The algorithm struggles on elections with large numbers of candidates,
essentially since the search space grows as the factorial of the number of
candidates,
but provides an important first step in tackling this important and
challenging problem.

As future work we plan to adapt the {\bf margin-stv} algorithm to answer
specific questions such as the influence of instances of multiple-voting on
election outcomes, and the impact of losing votes, similar to our prior
work~\citep{blom16} on  Instant Runoff Voting (IRV) elections. In addition, we aim to explore 
how applying the algorithm to the \textit{tail} of an STV election (i.e., by fixing the 
elections and eliminations that occur in the first $R$ rounds of counting) can be used within
a practical algorithm for computing improved lower and upper bounds on the MOV in challenging
instances (with a large number of seats or candidates).

\appendix
\section{US Election Data (2007 to 2010)}\label{sec:IRVlist}

In the experiments of Section \ref{sec:Evaluation} we reimagine a series of 16 IRV elections conducted in the US between 2007 and 2010 as STV elections with 2 available seats. Table \ref{tab:IRVdetails} records the name, number of candidates, and number of votes cast, for each of these 16 elections. Candidate numbers range between 4 and 13.

{\footnotesize
\begin{table}[t]
\centering
\begin{tabular}{|c|c|c|c|c}
\hline
\textbf{\#} & \textbf{Election Name} & \textbf{Candidates} & \textbf{Votes Cast} \\
\hline
1 & Pierce 2008 County Auditor & 4 & 159,987 \\
2 & Pierce 2008 City Council & 4 & 43,661 \\
3 & San Leandro 2010 Mayor & 4 & 23,494 \\
4 & Oakland 2010 D6 City Council & 4 & 14,040 \\
5 & Berkeley 2010 D8 City Council & 4 & 5,333 \\
6 & Berkeley 2010 D7 City Council & 4 & 4,862 \\  
\hline
7 & Pierce 2008 County Executive & 5 & 312,771 \\
8 & Berkeley 2010 D4 City Council & 5 & 5,708 \\ 
9 & Berkeley 2010 D1 City Council & 5 & 6,426 \\
10 & Aspen 2009 Mayor & 5 & 2,544 \\
\hline 
11 & Pierce 2008 County Assessor & 7 & 43,661 \\
12 & San Leandro 2010 D5 City Council & 7 & 23,494 \\
\hline
13 & Oakland 2010 D4 City Council & 8 &  23,884 \\
\hline
14 & Oakland 2010 Mayor & 11 & 122,268 \\
15 & Aspen 2009 City Council & 11 & 2,544 \\
\hline
16 & San Francisco 2007 Mayor & 13 & 149,465 \\
\hline
\end{tabular}
\caption{Name, number of candidates, and number of votes cast, for 16 IRV elections held in the US between 2007 and 2010, reimagined as STV elections with 2 available seats.}
\label{tab:IRVdetails}
\end{table}
}

\bibliographystyle{abbrvnat}
\bibliography{STVMargins}

\end{document}